\documentclass[12pt,onecolumn]{article}
\usepackage{amsfonts}
\usepackage{mathrsfs}
\usepackage[dvips]{graphicx}
\usepackage{multirow,tabularx}
\usepackage{subfigure}
\usepackage{amssymb}
\usepackage{amsmath}
\usepackage{amsbsy}
\usepackage{amsthm}
\usepackage{slashbox}

\usepackage{longtable}
\usepackage{algorithm}
\usepackage{algorithmic}
\usepackage{rotating}
\usepackage{textcomp,booktabs,threeparttable}
\usepackage[usenames,dvipsnames]{color}
\usepackage{colortbl}
\definecolor{mygray}{gray}{.9}
\definecolor{mypink}{rgb}{.99,.91,.95}
\definecolor{mycyan}{cmyk}{.3,0,0,0}
\definecolor{myred}{cmyk}{1,0,0,0}
\usepackage{url}

\def \bfW {\mathbf{W}}
\def \bfX {\mathbf{X}}

\def \bfy {\mathbf{y}}

\def \bfbeta {\boldsymbol{\beta}}

\def \bfA {\mathbf{A}}

\def \bfe {\mathbf{e}}

\def \bfI {\mathbf{I}}
\def \bfu {\mathbf{u}}

\def \pr {\mbox{Pr}}

\def\marginset#1#2{                      

\setlength{\oddsidemargin}{#1}         

\setlength{\evensidemargin}{0mm}       

\setlength{\hoffset}{\paperwidth}

\addtolength{\hoffset}{-\oddsidemargin}

\addtolength{\hoffset}{-\textwidth}

\addtolength{\hoffset}{-\evensidemargin}

\setlength{\hoffset}{0.5\hoffset}

\addtolength{\hoffset}{-1in}           

\setlength{\voffset}{-1in}             

\setlength{\topmargin}{\paperheight}

\addtolength{\topmargin}{-\headheight}

\addtolength{\topmargin}{-\headsep}

\addtolength{\topmargin}{-\textheight}

\addtolength{\topmargin}{-\footskip}

\addtolength{\topmargin}{#2}           

\setlength{\topmargin}{0.5\topmargin}

}
\marginset{0mm}{0mm}

\begin{document}
\title{\bf GPA: A statistical approach to prioritizing GWAS results by integrating pleiotropy information and annotation data}
\author{Dongjun Chung$^{1}$\footnote{These authors contributed equally to this work.}, Can Yang$^{1, 2}$\footnotemark[1], Cong Li$^{3}$, Joel Gelernter$^{2, 4, 5, 6}$ and Hongyu Zhao$^{1}$\footnote{Corresponding author}\\
\small{$^{1}$Department of Biostatistics, Yale School of Public Health}, \\ \small{New Haven, Connecticut,
USA.}\\
\small{$^{2}$Department of Psychiatry, Yale School of Medicine}, \\\small{New Haven, Connecticut, USA.} \\
\small{ $^3$Program in Computational Biology and Bioinformatics, Yale University,} \\ \small{New Haven, Connecticut, USA.}\\
\small{$^4$VA CT Healthcare Center, West Haven, Connecticut, USA.}\\
\small{$^5$Department of Genetics, Yale School of Medicine} \\ \small{West Haven, Connecticut, USA.}\\
\small{$^6$Department of Neurobiology, Yale School of Medicine} \\ \small{West Haven, Connecticut, USA.}
}

\date{\today} \maketitle

\begin{abstract}
Genome-wide association studies (GWAS) suggests that a complex disease is typically affected by many genetic variants with small or moderate effects.
Identification of these risk variants remains to be a very challenging problem.
Traditional approaches focusing on a single GWAS dataset alone ignore relevant information that could potentially improve our ability to detect these variants:
(1) Accumulating evidence suggests that different complex diseases are genetically correlated,
 i.e., multiple diseases share common risk genetic bases, which is known as ``pleiotropy''.
(2) SNPs are not equally important and functionally annotated genetic variants have demonstrated a consistent pattern of enrichment.
Thus, we proposed a novel statistical approach, named GPA, to performing integrative analysis of multiple GWAS datasets and functional annotations.
Hypothesis testing procedures were developed to facilitate statistical inference of pleiotropy and enrichment of functional annotation.
A computationally efficient EM algorithm was also available to handle millions of SNP markers.
Notably, our approach takes SNP-wise $p$ values from traditional GWAS analysis as input,
making it especially useful when only SNP summary statistics are available but not the genotype and phenotype data.
We applied our approach to perform systematic analysis of five psychiatric disorders. Not only did GPA identify many weak signals missed by the original single phenotype analysis, but also revealed interesting genetic architectures of these disorders.
the pleiotropic effect was detected among these among psychiatric disorders, and the SNPs in the central nervous system genes were significantly enriched.
These results deepened our understanding of genetic etiology for psychiatric disorders. We also applied GPA to the bladder cancer GWAS data with the ENCODE DNase-seq data from 125 cell lines and showed that GPA can detect cell lines that are more biologically  relevant to the phenotype of interest.
In summary, GPA can serve as an effective tool for integrative data analysis in the post-GWAS era.
\end{abstract}

\section{Introduction}

 Hundreds of genome-wide association studies (GWAS) have been conducted to study the genetic bases of complex human traits/diseases. As of Dec., 2013, more than 11,000 single-nucleotide polymorphisms (SNPs) have been reported to be significantly associated with at least one complex traits/diseases (see the web resource of GWAS catalog \cite{hindorff2009potential} \url{http://www.genome.gov/gwastudies/}). Despite of these successes of GWAS, these significantly associated SNPs can only explain a small portion of genetic contributions to complex traits/diseases \cite{manolio2009finding}. For example, human height is a highly heritable trait whose heritability is estimated to be around 80\%, i.e., 80\% of variation in height within the same population can be attributed to genetic effects \cite{visscher2008heritability}.  Based on large-scale GWAS, about 180 SNPs have been reported to be significantly associated with human height \cite{allen2010hundreds}. However, these loci together can only explain about 5-10\% of variation in height \cite{allen2010hundreds,manolio2009finding, visscher2008sizing}. This phenomenon is referred to as ``the missing heritability'' \cite{maher2008personal,Manolio2010,manolio2009finding}.

Finding the missing heritability has drawn much attention from worldwide researchers. The role of common variants (minor allele frequency (MAF) $\geq$ 0.01)
has been shown to be critical in explaining the phenotypic variance.
Instead of only using genome-wide significant SNPs, Yang et al. \cite{yang2010common} reported that, by using all genotyped common SNPs,
45\% of the variance for human height can be explained. This result suggests that a large proportion of
the heritability is not actually missing: given the limited sample size,
many individual effects of genetic markers are too weak to pass the genome-wide significance,
and thus those variants remain undiscovered. So far, people have found similar genetic architectures for many
other complex traits/diseases \cite{visscher2012five}, such as metabolic syndrome traits \cite{vattikuti2012heritability}
and psychiatric disorders \cite{pgc2013genetic,lee2012estimating}, i.e.,
the phenotype is affected by many genetic variants with small or modest effects,
which is usually referred to as ``polygenicity''. The polygenicity of complex traits is further
supported by recent GWAS with larger sample sizes, in which more associated common SNPs with moderate
 effects have been identified (e.g., \cite{morris2012large}). Clearly, the emerging polygenic genetic architecture
 imposes a great challenge of identifying risk genetic variants: a larger sample size is required to identify
 genetic variants with smaller effect sizes. However, sample recruitment may be expensive and time-consuming.
 Alternatively, integrative analysis of genomic data could be a promising direction, including combining GWAS data of multiple genetically related phenotypes
 and incorporating relevant biological information.

The last couple of years have seen an emerging recognition of ``pleiotropy'', i.e. the sharing of genetic factors,
between human complex traits/diseases. For example, a systematic analysis of the NHGRI catalog (\url{http://www.genome.gov/gwastudies/})
of published GWAS showed that 16.9\% of the reported genes and 4.6\% of the reported SNPs
are associated with multiple traits/diseases \cite{sivakumaran2011abundant}.
Through a ``pleiotropic enrichment'' method, Andreassen et al showed that it is possible to improve the power to
 detect schizophrenia-associated genetic variants by utilizing the pleiotropy between schizophrenia (SCZ)
 and cardiovascular-disease \cite{andreassen2013improved}. A more recent study identified four significant loci
  ($p$-value $< 5\times10^{-8}$) to be pleiotropic by analyzing GWAS data of 33,332 cases and 27,888 controls for
  five psychiatric disorders \cite{pgp2013improved}. Further analysis suggested very significant genetic correlation
  between schizophrenia and bipolar disorder ($0.68\pm0.04$ s.e.) \cite{pgc2013genetic}.
  The wide-existing pleiotropy has also been demonstrated among several other types of traits, for example,
   metabolic syndrome traits \cite{vattikuti2012heritability} and cancers \cite{sakoda2013turning}.

An increasing number of studies also suggest that functionally annotated SNPs are generally more biologically important and henceforth more likely to be associated with complex traits/diseases. To name a few, Schork et al. \cite{schork2013all} demonstrated a consistent pattern of enrichment of GWAS signals among functionally annotated SNPs, using GWAS data of different traits/diseases (e.g., Crohn's disease and SCZ). Yang et al. \cite{yang2011genome} showed that SNPs in the genic region could explain more variance of height and body mass index (BMI) than SNPs in the intergenic region. Nicolae et al. \cite{nicolae2010trait} suggested that SNPs associated with complex traits/disease were more likely to be expression quantitative trait loci (eQTL). In addition, public availability of the vast amount of functional annotation data also provides unprecedented opportunities to investigate
 the enrichment of GWAS signals among these various types of functional annotations. For example, recently, the Encyclopedia of DNA Elements (ENCODE) Consortium has generated vast amounts of experimental data on gene expression (RNA-seq), DNA methylation status (RRBS-seq), chromatin modifications (ChIP-seq), chromatin accessibility (DNase-seq and FAIRE-seq), transcription factor (TF) binding sites (ChIP-seq), and long-range chromatin interactions (ChIA-PET, Hi-C, and 5C). As of September 2012, more than 1,600 data sets from 147 cell lines have been produced to annotate human genome, including 2.89 million unique, non-overlapping DNase I hypersensitivity sites (DHSs) in 125 cell lines using DNase-seq and 630K binding regions of 119 DNA-binding proteins in 72 cell lines using ChIP-seq, among many \cite{encode2012}. The ENCODE Project Consortium \cite{encode2012} examined 4,492 risk-associated SNPs from NHGRI GWAS catalog and found that 12\% of them overlap with TF binding regions and 34\% overlap with DHSs.

The widely existing pleiotropy between complex traits/diseases and increasingly rich functional annotation data are calling for novel statistical
 methods to effectively perform joint analysis of multiple GWAS data sets and functional annotation data. Statistical methods to investigate pleiotropy have
 been actively researched (reviewed in \cite{solovieff2013pleiotropy} and \cite{shriner2012moving}), for example, using linear mixed models \cite{lee2012estimation, li2013improving}. However, these methods do not allow to utilize functional annotation data for  prioritization of GWAS results.
 On the other hand, various statistical methods which make use of functionally annotated SNPs have been proposed in recent years (reviewed in \cite{edwards2013beyond}, \cite{cantor2010}, and \cite{noncoding2012}). For example, GSEA \cite{subramanian2005gene} identifies potentially important pathways in which target genes of risk-associated SNPs are involved while RegulomeDB \cite{regulomedb} allows nucleotide-level annotations of risk-associated SNPs, especially for those located in non-coding regions. Stratified False Discovery Rate methods were applied
 to incorporate annotation into GWAS data analysis \cite{schork2013all}. However, these methods are designed for the analysis of single phenotype and hence, these methods do not utilize functional annotation data fully efficiently for the genetic variants shared by multiple phenotypes.
In short, well established statistical framework for the integration of functional annotation data to the joint analysis of genetically correlated GWAS data is still missing in the current literature.

In order to meet the emerging needs described above, we propose a unified statistical framework, named GPA
\footnote{``G'' for GWAS, ``P'' for pleiotropy and ``A'' for Annotation.}, to prioritize GWAS result based on the pleitropy and annotation information.
GPA also provides statistically rigorous and easily interpretable inference tools for the analysis of pleiotropy and the enrichment of functional annotation. This article is organized as follows. In Section 2, we first describe our GPA probabilistic model and its statistical inference procedures. In Section 3, we investigate the properties of GPA using extensive simulation studies and illustrate the versatility and utility of GPA with the analysis of real data. Specifically, we apply GPA to the five psychiatric disorder GWAS data with central nervous system gene expression data and show that GPA can accurately identify pleiotropy structure among these diseases. We further apply GPA to the bladder cancer GWAS data with the ENCODE DNase-seq data from 125 cell lines and show that GPA can detect cell lines that are biologically more relevant to the phenotype of interest. In Section 4, we conclude this paper with the discussion of related issues.

\section{Methods}


\subsection{GPA probabilistic model}

Throughout this paper, we shall use $j$ to index SNPs, $k$ to index GWAS data sets, $d$ to index the annotation data sets. Let's first consider the simplest case that we only have summary statistics ($p$-values) from only one GWAS data set, then extend our model to handle multiple GWAS data sets and annotation data. Suppose we have performed hypothesis testing of genome-wide SNPs and obtained their $p$-values:
  \begin{equation}\label{}
   \begin{aligned}
   \text{Null hypothesis}: & \quad H^{(1)}_0,H^{(2)}_0,\dots,H^{(j)}_0,\dots,H^{(M)}_0,\\
   \text{$p$-value}: & \quad P_1,P_2,\dots,P_j,\dots,P_M,
   \end{aligned}
  \end{equation}
where $M$ is the number of SNPs. Consider the ``two-groups model'' \cite{efron2008microarrays}, i.e., the obtained $p$-values are assumed to come from the mixture of null and non-null, with probability $\pi_0$ and $\pi_1 = 1-\pi_0$, respectively. Let $\mathbf{Z}_{j}=[Z_{j0},Z_{j1}]$ be the latent variables indicating whether the $j$-th SNP is null or non-null, where $Z_{j0}\in \{0,1\}$, $Z_{j1}\in \{0,1\}$, and $Z_{j0} + Z_{j1} = 1$ (A SNP can only be either null or non-null). $Z_{j0}=1$ means un-associated (null) and $Z_{j1}=1$ means associated (non-null). Then we have the following two-groups model:
   \begin{equation}\label{twogroup2}
   \begin{aligned}
   \pi_0 = \pr(Z_{j0}=1): & \quad ( P_j | Z_{j0}=1 ) \sim \mathcal{U}[0,1], \\
   \pi_1 = \pr(Z_{j1}=1): & \quad ( P_j | Z_{j1}=1 ) \sim Beta(\alpha,1).
   \end{aligned}
  \end{equation}
where the $p$-values from the null group are from the Uniform distribution on [0,1], denoted as $\mathcal{U}[0,1]$, and the $p$-values from the non-null group are from the Beta distribution with parameters ($\alpha,1$), where $0<\alpha<1$. We put the constraint $0<\alpha<1$ to model that smaller $p$-value is more likely than larger $p$-value when it is from the non-null group \cite{pounds2003}.

To incorporate information from functional annotation data, we extend the basic model as follows. Suppose we have collected information from $D$ functional annotation sources in the annotation matrix: $\mathbf{A}\in \mathbb{R}^{M\times D}$, where $A_{j,d}\in\{0,1\}$ indicates whether the $j$-th SNP is annotated in the $d$-th functional annotation source. For example, when there are two annotation sources: eQTL data and DNase I hypersensitivity sites (DHS) data, then $\mathbf{A}$ is an $M\times 2$ matrix. If the $j$-th SNP is an eQTL, then $A_{j1}=1$, otherwise $A_{j1}=0$; if it is located in a DHS, then $A_{j2}=1$, otherwise $A_{j2}=0$. Now we model the relationship between $\mathbf{Z}_j$ and $A_{jd}$ as
\begin{align}\label{fourgroup2}
	(A_{jd}|Z_{j0}=1) \sim Bernoulli( q_{d0} ), \nonumber\\
	(A_{jd}|Z_{j1}=1) \sim Bernoulli( q_{d1} ).
\end{align}
Clearly, $q_{d0}=\pr(A_{jd}=1|Z_{j0}=1)$ can be interpreted as the proportion of null SNPs in the $d$-th annotation file, and $q_{d1}=\pr(A_{jd}=1|Z_{j1}=1)$ corresponds to the proportion of non-null SNPs in the $d$-th annotation file. Therefore, $q_{d1}>>q_{d0}$ means that there exists enrichment in the $d$-th annotation file. The statistical inference about enrichment of annotation data will be discussed in details in Section \ref{section_pleiotropy_enrichment}.

Now we extend the above model to handle multiple GWAS data sets. To keep the notation uncluttered, we present the model for the case of two GWAS data but the generalization to more than two GWAS data is straightforward. In fact, our GPA model and its algorithm are not limited by the number of GWAS data. Suppose we have $p$-values from two genetically related GWAS:
  \begin{align}
   \text{$p$-value from GWAS1}: & P_{11},P_{21},\dots,P_{j1},\dots,P_{M1}.\nonumber\\
   \text{$p$-value from GWAS2}: & P_{12},P_{22},\dots,P_{j2},\dots,P_{M2}.
  \end{align}
Let $\mathbf{P}\in \mathbb{R}^{M\times 2}$ be the matrix collecting all the $p$-values, where $P_{jk}$ denotes the $p$-value of the $j$-th SNP in the $k$-th GWAS. Similarly, we introduce latent variables $\mathbf{Z}_j = [Z_{j00},Z_{j10},Z_{j01},Z_{j11}]$ indicating the association between the $j$-th SNP and  the two phenotypes: $Z_{j00}=1$ means the $j$-th SNP is associated with neither of them, $Z_{j,10}=1$ means it is only associated with the first one, $Z_{j01}=1$ means it is associated with the second one, and $Z_{j11}=1$ means it is associated with both.
The two-groups model (\ref{twogroup2}) is extended to the following ``four-groups model'':
   \begin{align}\label{fourgroup}
   \pi_{00} = \pr(Z_{j00}=1): & \quad ( P_{j1} | Z_{j00}=1 ) \sim \mathcal{U}[0,1],\quad \quad \,\,( P_{j2} | Z_{j00}=1 )\sim\mathcal{U}[0,1], \nonumber\\
   \pi_{10} = \pr(Z_{j10}=1): & \quad ( P_{j1} | Z_{j10}=1 )\sim Beta(\alpha_1,1)\,,( P_{j2} | Z_{j10}=1 )\sim\mathcal{U}[0,1], \nonumber\\
   \pi_{01} = \pr(Z_{j01}=1): & \quad ( P_{j1} | Z_{j01}=1 )\sim \mathcal{U}[0,1],\quad \quad \,\, ( P_{j2} | Z_{j01}=1 )\sim Beta(\alpha_2,1), \nonumber\\
   \pi_{11} = \pr(Z_{j11}=1): & \quad ( P_{j1} | Z_{j11}=1 )\sim Beta(\alpha_1,1)\,,( P_{j2} | Z_{j11}=1 )\sim Beta(\alpha_2,1),
  \end{align}
where $0<\alpha_k<1,k=1,2$. When the genetic basis of the two phenotypes are independent of each other (i.e., no pleiotropy), then we have $\pi_{11} = ( \pi_{10} + \pi_{11} ) ( \pi_{01} + \pi_{11} )$ by expectation. Therefore, the difference between $\pi_{11} $ and $ ( \pi_{10} + \pi_{11} ) ( \pi_{01} + \pi_{11} )$ can be used to characterize pleiotropy. Statistical inference on pleiotropy is given in Section \ref{section_pleiotropy_enrichment}.

To incorporate annotation information into multiple GWAS model (\ref{fourgroup}), similarly, we model the relationship between $\mathbf{Z}_j$ and $A_{jd}$ as
\begin{align}\label{fourgroup2}
	(A_{jd}|Z_{j00}=1) \sim Bernoulli( q_{d00} ), \nonumber\\
	(A_{jd}|Z_{j10}=1) \sim Bernoulli( q_{d10} ), \nonumber\\
	(A_{jd}|Z_{j01}=1) \sim Bernoulli( q_{d01} ), \nonumber\\
	(A_{jd}|Z_{j11}=1) \sim Bernoulli( q_{d11} ),
\end{align}
where $q_{d00}$ is the probability of a null SNP being annotated, $q_{d10}$ is the probability of the first phenotype associated-SNP  being annotated, $q_{d01}$ is the probability of the second phenotype associated-SNP  being annotated and $q_{d11}$ is the probability of jointly associated-SNP being annotated. Assuming the independence of SNP markers,  the joint distribution $\pr(\mathbf{P},\mathbf{A})$ can be written as
\begin{align}\label{GPA_model}
\pr(\mathbf{P},\mathbf{A}) &= \prod^M_{j=1} \left[\sum_{l\in\{00,10,01,11\}}\pr(Z_{jl}=1)\pr(\mathbf{P}_j,\mathbf{A}_j|Z_{jl}=1)\right]\nonumber\\
&=\prod^M_{j=1}\left[\sum_{l\in\{00,10,01,11\}}\pi_{l}\pr(\mathbf{P}_j|Z_{jl}=1)\pr(\mathbf{A}_j|Z_{jl}=1)\right]\nonumber\\
&=\prod^M_{j=1}\left[\sum_{l\in\{00,10,01,11\}}\pi_{l}\pr(\mathbf{P}_j|Z_{jl}=1)\prod^D_{d=1}{\pr({A}_{jd}|Z_{jl}=1)}\right],
\end{align}
where $\mathbf{P}_j$ and $\mathbf{A}_j$ are the $j$-th row of $\mathbf{P}$ and $\mathbf{A}$; the second equation holds by assuming the independence between $\mathbf{P}_j$ and $\mathbf{A}_j$, conditional on $Z_{jl}$; the third equation holds by further assuming the independence between $A_{jd}$ and $A_{j d'}$ for $d \neq d'$, conditional on ${Z}_{jl}$.

Parameters of the GPA probabilistic model are estimated using the Expectation-Maximization (EM) algorithm \cite{dempster1977},
which turns out to be computationally efficient because we have explicit solutions for estimation of all the parameters in the M-step. Standard errors for parameter estimates are estimated using the empirical observed information matrix \cite{mclachlan2008}. Note that in the GPA model, the sample size for estimating the empirical observed information matrix corresponds to the number of SNPs and as a result, we have a very large sample size ($\sim 10^6$) to accurately estimate standard errors. More details of the EM algorithm and the estimation of standard errors are provided in Appendix.

\subsection{Statistical inference}\label{section_inference}

\subsubsection{False discovery rate}

After we estimate parameters for the GPA model, SNPs can be prioritized based on their local false discovery rates \cite{efron2010large}. Note that separate analysis of single GWAS data based on their summary statistics is equivalent to the analysis of single GWAS data using GPA without any annotation data. Hence, separate analysis of single GWAS data can be considered as a special case of our GPA framework.

To present the local false discovery rate based on our GPA model, we begin with the simplest case: single GWAS without annotation data, In this case, there are only two groups: null and non-null. The false discovery rate is defined as the probability that $j$-th SNP belongs to the null group given its $p$-value, i.e.,
\begin{align}\label{fdr1}
\mbox{fdr}(P_j) = \pr(Z_{j0}=1|P_j).
\end{align}
For joint analysis of two GWAS data sets, we are interested in the local false discovery rate of the $j$-th SNP, if it is claimed to be associated with the first phenotype and the second one, i.e.,
\begin{align}\label{fdr2}
  &\mbox{fdr}_1(P_{j1},P_{j2})=\pr(Z_{j00}+Z_{j01}=1|P_{j1},P_{j2}),\nonumber\\
	&\mbox{fdr}_2(P_{j1},P_{j2})=\pr(Z_{j00}+Z_{j10}=1|P_{j1},P_{j2}).
\end{align}
Similarly, when annotation data are available, the false discovery rates can be calculated  as
\begin{align}\label{fdr3}
  &\mbox{fdr}_1(P_{j1},P_{j2},\bfA)=\pr(Z_{j00}+Z_{j01}=1|P_{j1},P_{j2},\bfA),\nonumber\\
	&\mbox{fdr}_2(P_{j1},P_{j2},\bfA)=\pr(Z_{j00}+Z_{j10}=1|P_{j1},P_{j2},\bfA).
\end{align}
Then, we use the \textit{direct posterior probability approach} \cite{newton2004} to control global false discovery rates. More details for the estimation of false discovery rates are provided in Appendix.

\subsubsection{Hypothesis testing of annotation enrichment and pleiotropy}\label{section_pleiotropy_enrichment}

Consider a given annotation file, we are particularly interested in the significance of its enrichment.
First, we propose to the likelihood ratio test (LRT) to assess the significance. To keep the notation simple, we drop the index of annotation files. Specifically, the significance of enrichment of an annotation file for single-GWAS data analysis can be assessed by the following hypothesis:
\begin{equation}\label{hypotheses0}
H_0: q_{0} = q_{1} \mbox{ v.s. } H_1: q_{0}\neq q_{1}.
\end{equation}
The LRT statistics is given as follows:
\begin{eqnarray*}
\lambda^{(A)} = \frac{ \pr( \mathbf{P},\mathbf{A} ; \hat \Theta^{(A)}_0 ) }{ \pr( \mathbf{P},\mathbf{A} ; \hat \Theta ) },
\end{eqnarray*}
where $\hat \Theta^{(A)}_0$ is the parameter estimates obtained under $H_0$ (Here the superscript \emph{A} indicates the Annotation enrichment test). Note that $\hat \Theta^{(A)}_0$ can be easily obtained by running the GPA algorithm without incorporating the annotation data. Under the null, the test statistic $-2 \log \lambda^{(A)}$ asymptotically follows the $\chi^2$ distribution with degree of freedom $df=1$ \cite{shao2003}. We reject $H_0$ if $-2 \log \lambda^{(A)} > \chi_{ df=1, \alpha }^2$, where $\chi_{ df=1, \alpha }^2$ is the $( 1 - \alpha )$-th quantile of $\chi^2$ distribution with $df=1$.

For joint analysis of two GWAS with annotation data, hypotheses (\ref{hypotheses0}) become as
\begin{equation}\label{joint-ann-hypo}
  H_0: q_{00} =  q_{10}= q_{01} = q_{11} \mbox{ v.s. } H_1: \mbox{ not } H_0.
\end{equation}
Under the null, the test statistics asymptotically follows $\chi^2$ distribution with $df=3$. Similarly, the test of annotation enrichment can be extended to handle $K$ GWAS. In this case, the test statistics asymptotically follows $\chi^2$ distribution with $df=2^K-1$ under the null.

Now we consider testing pleiotropy between two GWAS. When there is no pleiotropy, i.e., the signals from the two GWAS are independent of each other, testing pleiotropy can be formulated by evaluating the following hypothesis:
\begin{equation}\label{hypotheses2}
H_0:  \pi_{11} = \pi_{1*} \pi_{* 1},\mbox{ v.s. } H_1: \mbox{ not } H_0,
\end{equation}
where $\pi_{1*}=\pi_{10 }+\pi_{11}$, $\pi_{* 1}=\pi_{0 1}+\pi_{11}$.  The LRT statistics is constructed as follows:
\begin{eqnarray*}
\lambda^{(P)} = \frac{ \pr( \mathbf{P},\mathbf{A} ; \hat \Theta^{(P)}_0 ) }{ \pr( \mathbf{P},\mathbf{A} ; \hat \Theta ) },
\end{eqnarray*}
where $\hat \Theta^{(P)}_0$ is the parameter estimates obtained under $H_0$ (Here the superscript \emph{P} indicates the Pleiotropy test). The test statistics ($-2\log\lambda^{(P)}$) asymptotically follows $\chi^2$ distribution with $df=1$ under the null.

%

\section{Results}

\subsection{Simulation study}

We first evaluated performance of GPA for parameter estimation and hypothesis testing and compared them with some related methods using simulation studies.
We followed the classical liability threshold model \cite{lee2012estimating} to simulate case-control GWAS data for two genetically correlated diseases. For each disease, we first simulated a large cohort of individuals with genotypes of $M$ independent SNPs. The MAFs of these SNPs were drawn uniformly from [0.05, 0.5]. Then we randomly chose $m$ SNPs to be causal SNPs. The per-minor-allele effect of each causal SNP was drawn from a normal distribution with zero-mean and variance of $\frac{h^2}{(1-h^2)f_j(1-f_j)m}$, where $h^2$ is the desired level of variance explained by all SNPs on the liability scale and $f_i$ is the MAF of the corresponding causal SNP. We also simulated the the environmental effect on the liability scale for each individual from a standard normal distribution (zero mean and unit variance). The total liability for each individual was then obtained by adding up all the genetic effects and the environmental effect. Given a desired disease prevalence $B$, individuals with liabilities greater than the $1-B$ quantile were classified as cases and others were classified as controls. Then equal numbers of cases and controls were drawn from the cohort as a GWAS data set. When simulating two diseases simultaneously, we simulated two disjoint cohorts with the same set of SNPs. To reflect the pleiotropy effects between the two diseases, $m'$ causal SNPs ($m' \leq m$) were chosen to be shared by the two diseases. The annotation status of each causal and non-causal SNP was simulated from a Bernoulli distribution with probability of $q_1$ and $q_0$ respectively.

In our simulation study, the sample size of each data set, $N$, was set at 1000, 2000, 5000 or 10000. The number of causal SNPs $m$ was the same for the two diseases and was set at 500, 1000 or 2000. We varied the proportion of shared causal SNPs between the two diseases, $\gamma$, from $m/M$ to 1. Note that $\gamma = m/M$ corresponds to zero pleiotropy. The disease prevalence, $B$, was fixed at 0.1 and the variance explained by all the SNPs, $h^2$, was fixed at 0.6 for the two diseases. Here $q_1$ and $q_0$ were fixed at 0.4 and 0.1, respectively.

\begin{figure}
  \centering
  \includegraphics[width=.9\linewidth]{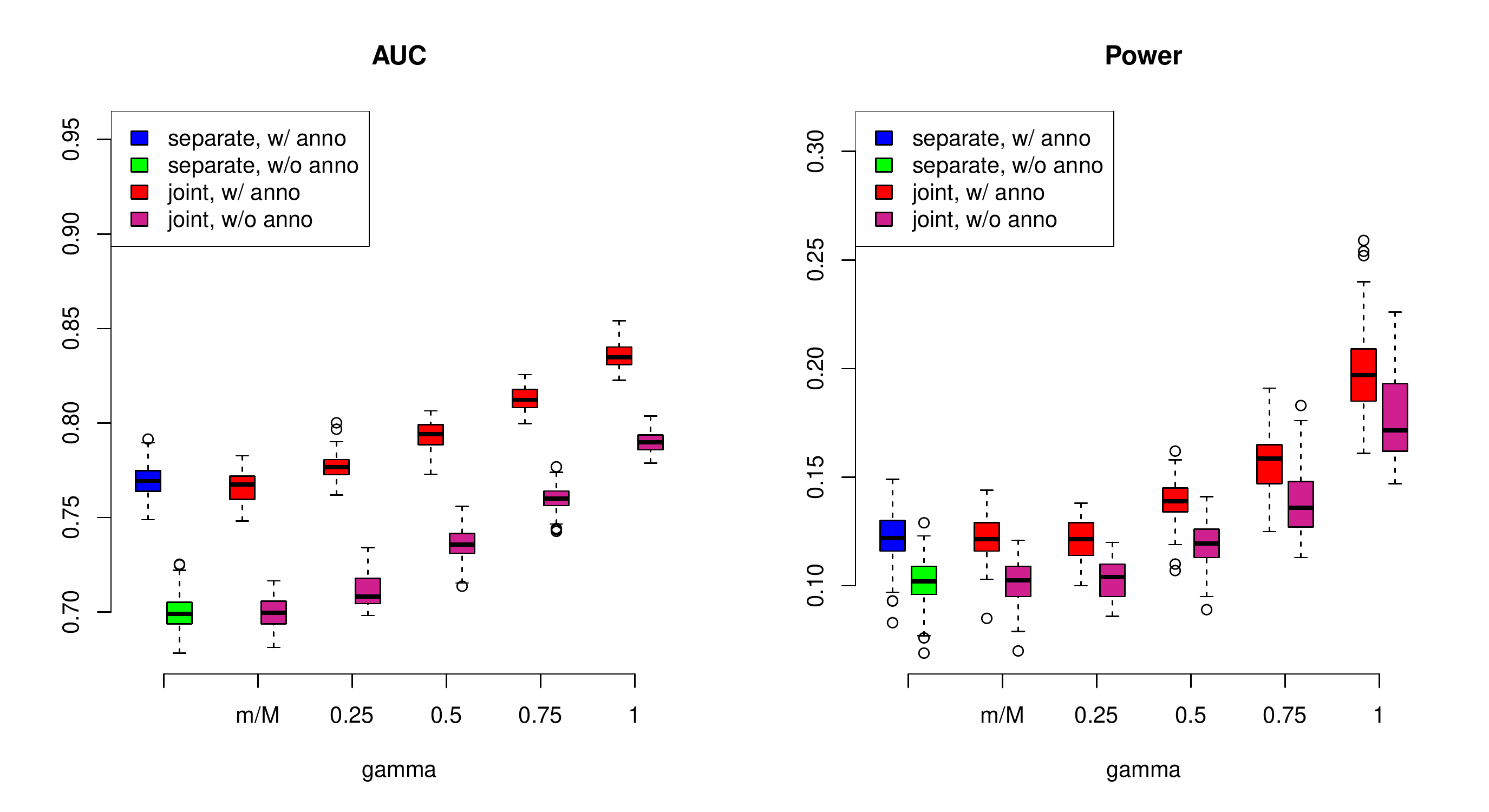}
  \caption{The AUC (left) and power (right) of GPA for SNP prioritization accuracy at $N$ = 5000, $m$ = 1000.}\label{AUC_Power}
\end{figure}

\begin{figure}
  \centering
  \includegraphics[width=.9\linewidth]{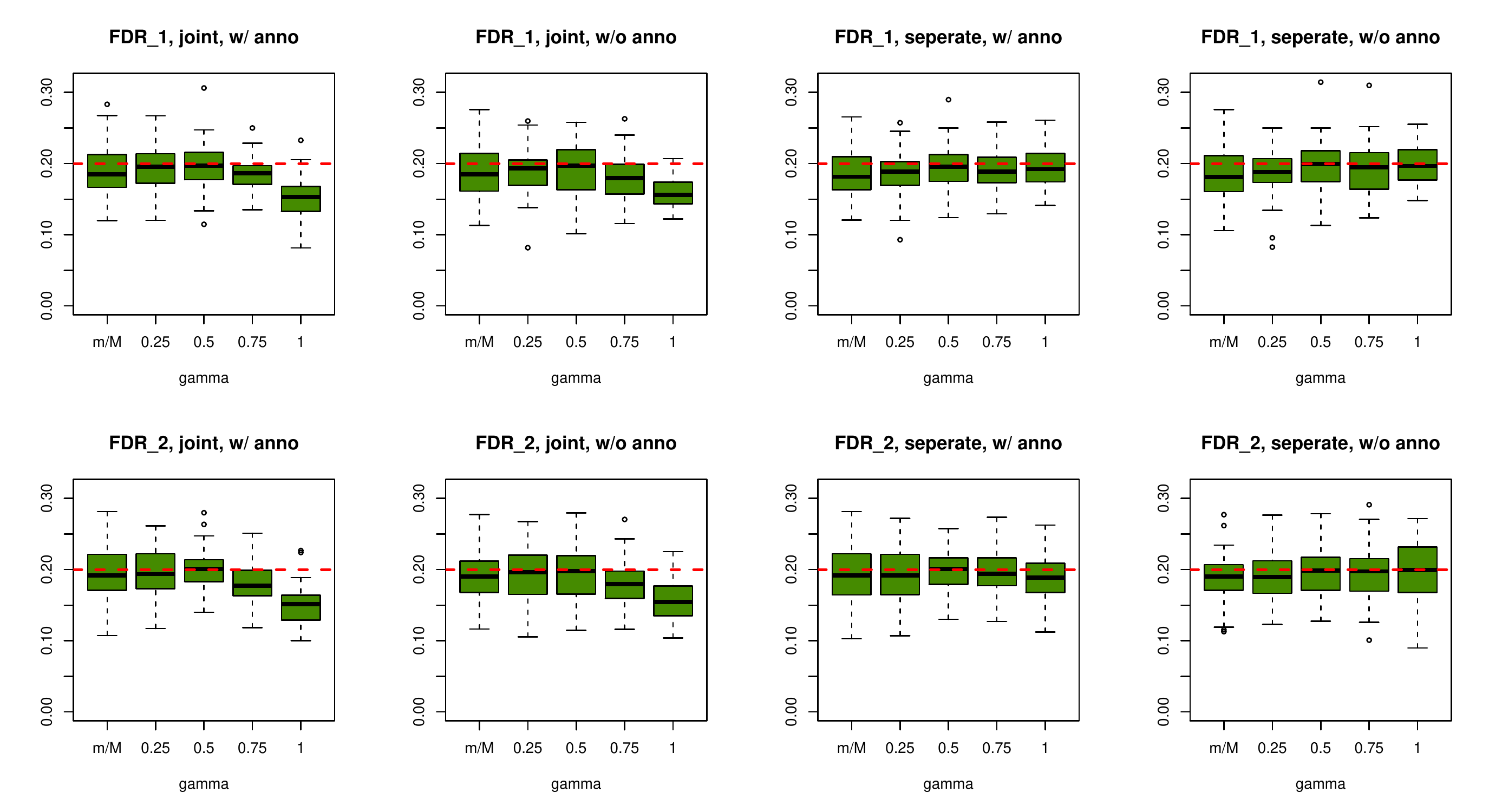}
  \caption{The global false discovery rates of GPA at $N$ = 5000, $m$ = 1000.
   The upper and lower panels show the results (joint and separate analysis, with and without annotation) for the first and second GWAS , respectively.
   For all scenarios, the global false discovery rates of GPA are controlled at the nominal level. }\label{FDR}
\end{figure}

\begin{figure}
  \centering
  \includegraphics[width=.9\linewidth]{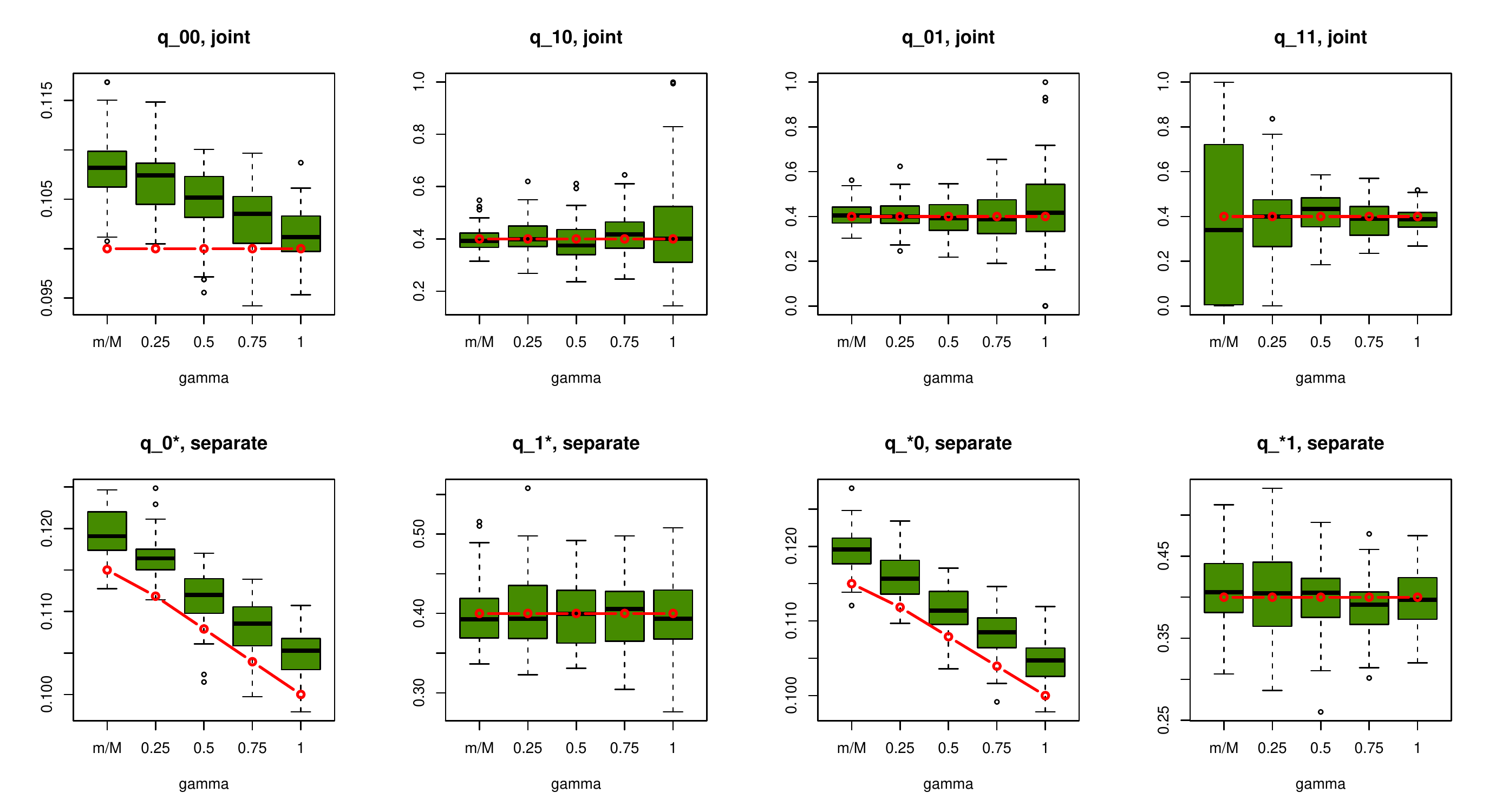}
  \caption{The estimated $q$ at $N$ = 5000, $m$ = 1000. The upper panel shows the estimated $q_{00},q_{10},q_{01},q_{11}$ in the joint analysis of two GWAS.
  The lower panel show the estimated $q_{0*},q_{1*}$ and $q_{*0},q_{*1}$ in the separate analysis for the first and second GWAS, respectively.
  The red lines represent the true values.}\label{Q}
\end{figure}

\begin{figure}
  \centering
  \includegraphics[width=.9\linewidth]{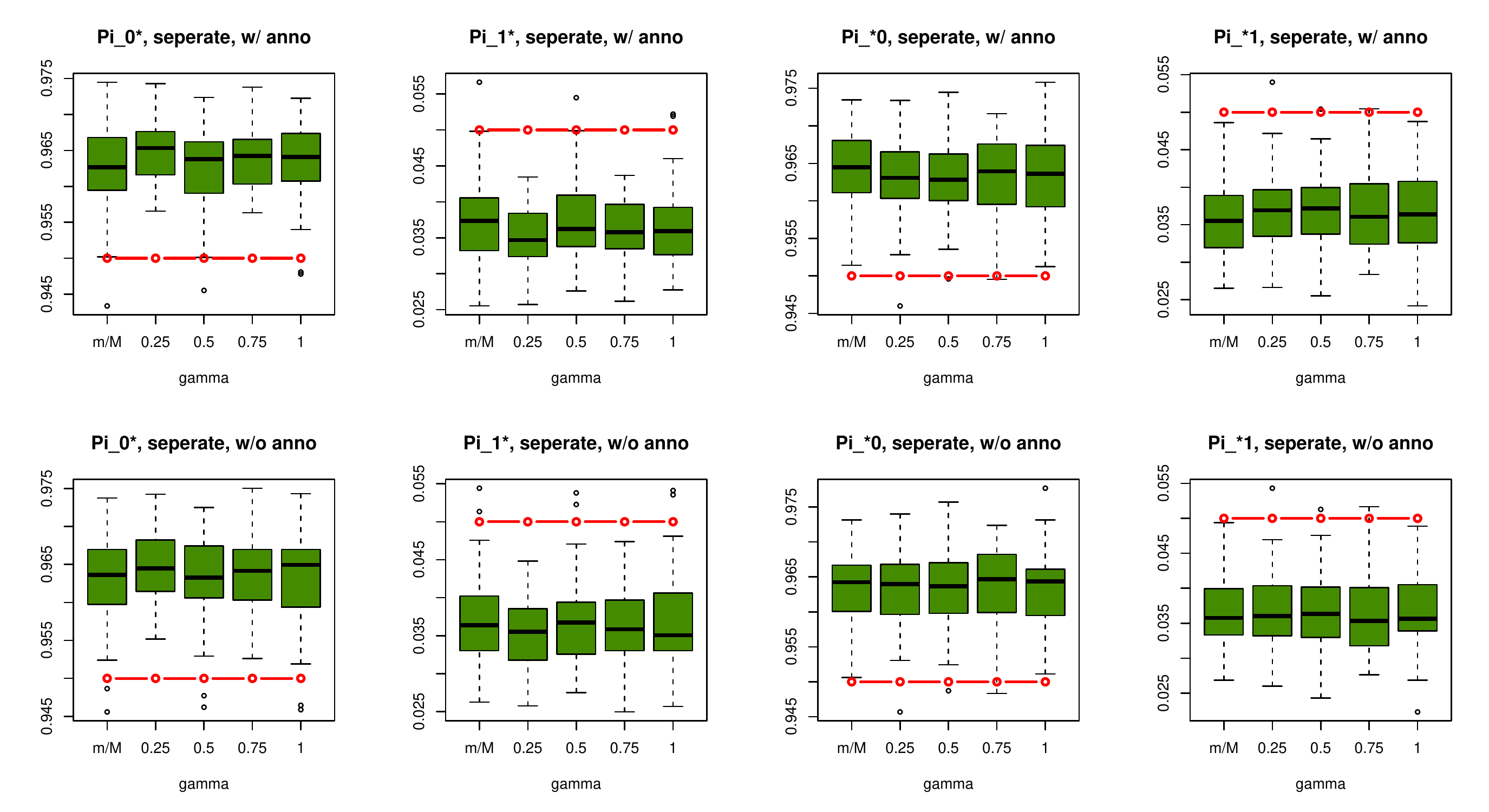}
  \caption{The estimated $\pi$ of separate analysis (single GWAS) at $N$ = 5000, $m$ = 1000. The upper panel shows the estimated $\pi_{0*},\pi_{1*},\pi_{*0},\pi_{*1}$ with annotation,
  and the lower panel shows the estimated $\pi_{0*},\pi_{1*}$ and $\pi_{*0},\pi_{*1}$ without annotation. The red lines represent the true values.}\label{Pi_marg}
\end{figure}

\begin{figure}
  \centering
  \includegraphics[width=.9\linewidth]{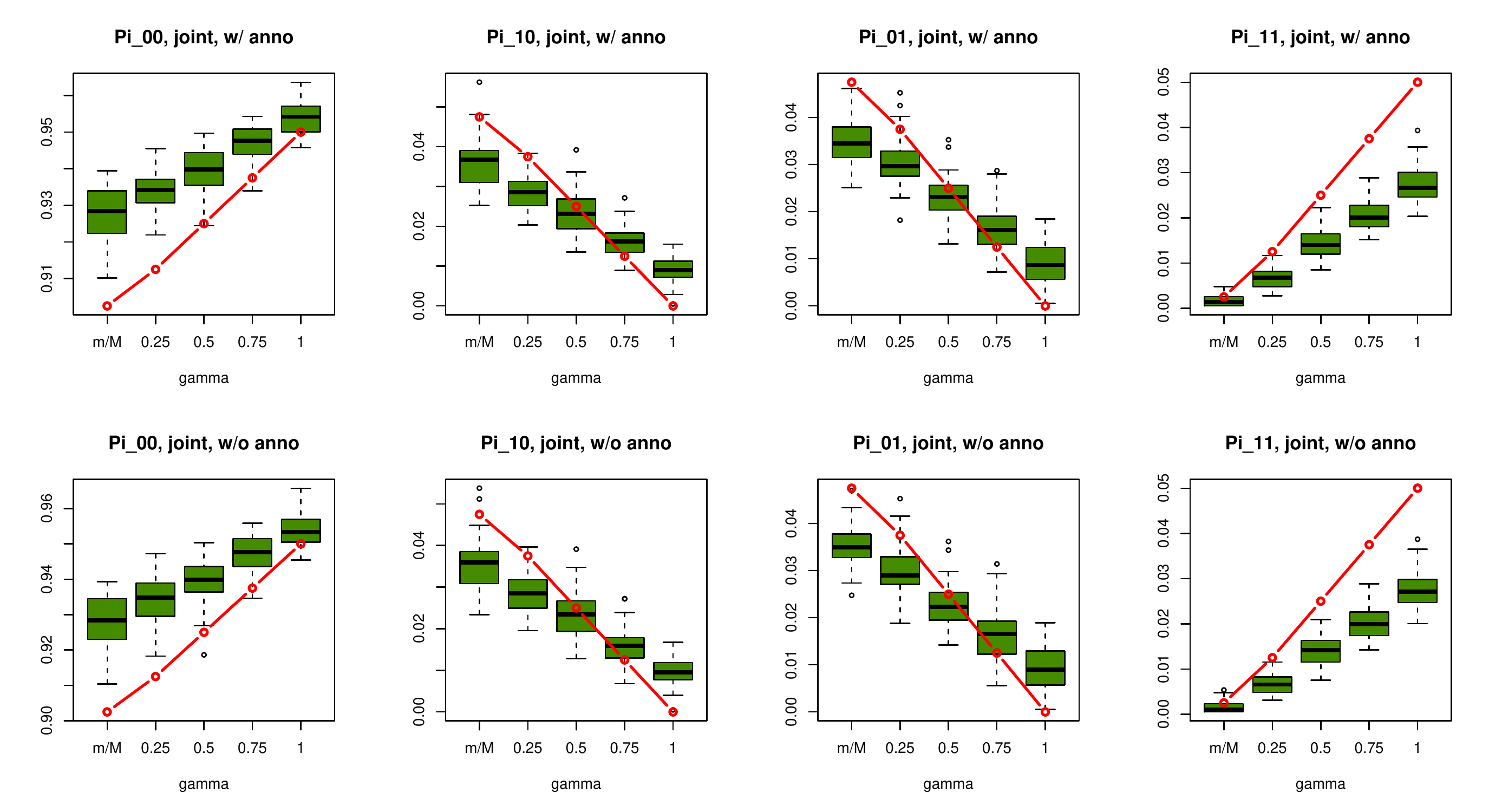}
  \caption{The estimated $\pi$ of joint analysis (two GWAS) at $N$ = 5000, $m$ = 1000. The upper panel shows the estimated $\pi_{00},\pi_{10},\pi_{01},\pi_{11}$ with annotation,
  and the lower panel shows the estimated $\pi_{00},\pi_{10},\pi_{01},\pi_{11}$ without annotation. The red lines represent the true values.}\label{Pi_joint}
\end{figure}

We first evaluated SNP prioritization accuracy of GPA. Specifically, after the two data sets were simulated, we obtained the $p$-value for each SNP in each disease using a one degree of freedom $\chi^2$ test. Then we analyzed the simulated data using our GPA method in the following four modes: 1. analyzing the two diseases separately without the annotation data; 2. analyzing the two diseases with the annotation data; 3. analyzing the two diseases jointly without the annotation data; 4. analyzing the two diseases jointly with the annotation data. The simulation was repeated 50 times for each scenario. In each mode, we compared the order of the local false discovery rates obtained using GPA against the actual causality status of the SNPs to calculate the area under the \emph{receiver operating characteristic} curve (AUC) as a measure of causal SNP prioritization accuracy. Figure \ref{AUC_Power} (left panel) shows the AUCs from the four modes with $N = 5000$ and $m = 1000$ (results for other scenarios are in the supplementary material). Because all the simulation parameters are the same for the two diseases, only the results for the first disease is shown. The results suggest that both the annotation information and the pleiotropy between the two diseases improved the prioritization accuracy. In particular, as the fraction of shared causal SNPs increased, the prioritization accuracy also increased. Given the local false discovery rates obtained using GPA, we controlled the global false discovery rate at 0.2 and calculated the average power for all the true causal SNPs. The results for $N = 5000$ and $m = 1000$ are shown in Figure \ref{AUC_Power} (right panel) and are similar with results of the AUC (results for other scenarios are in the supplementary material, Figure S1-S11). We also checked the actual false discovery rates and found that on average the false discovery rate was indeed by controlled at 0.2 in spite of occasional slight conservativeness (Figure \ref{FDR}, Figure S12-S22). In the simulations, we also found that GPA provided satisfactory estimate of $q$, the probability of being annotated for a certain group of SNPs,
 as long as there are enough number of SNPs in that group (Figure \ref{Q}, Figure S23-S33). However, we note that the proportion of causal SNPs always tends to be under-estimated (Figure \ref{Pi_marg}, Figure \ref{Pi_joint}, Figure S34-S55). The under-estimation may be due to the limited sample size and the model mismatch
between GPA and the random-effects model.


\begin{figure}
  \centering
  \includegraphics[width=.5\linewidth]{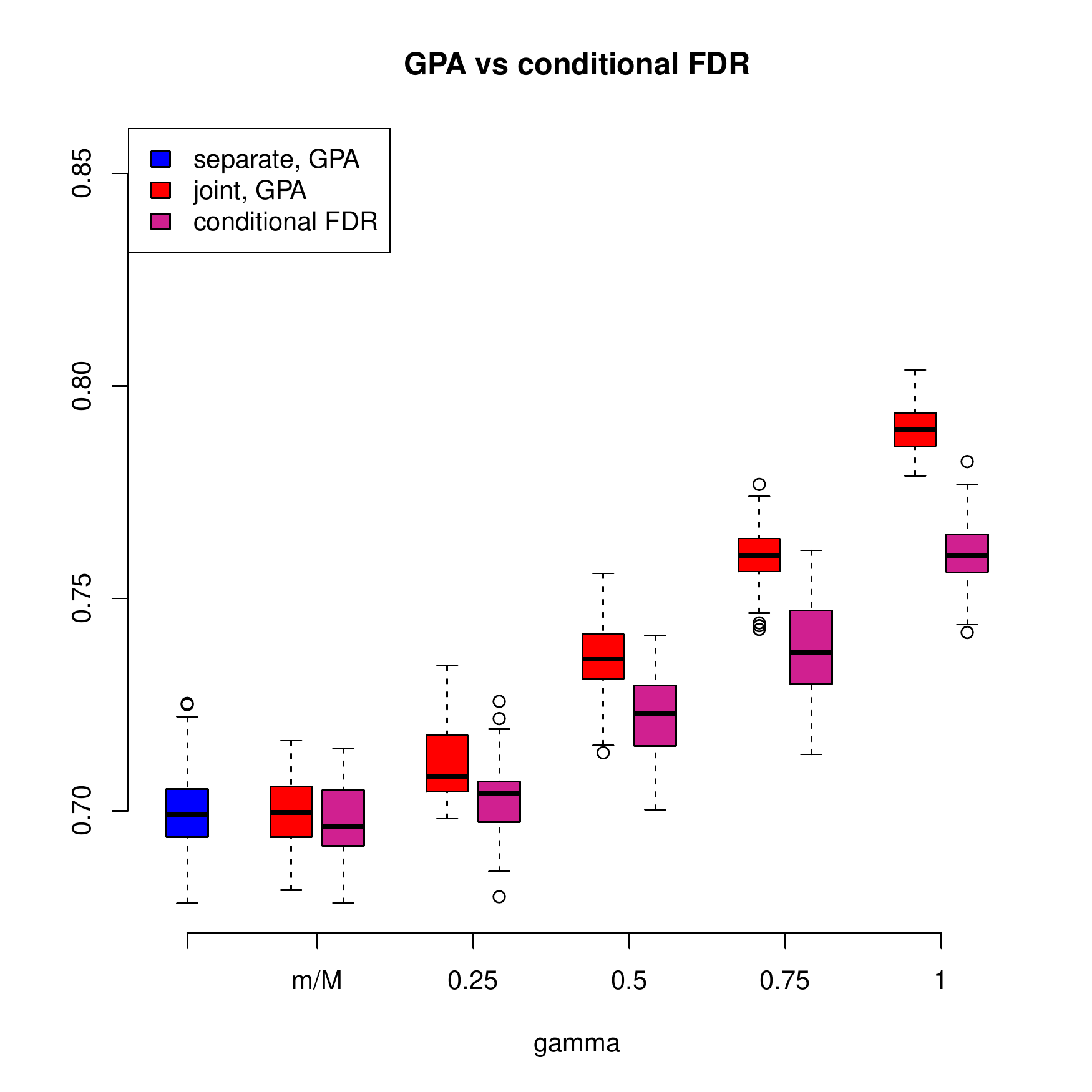}
  \caption{The comparison of AUC between GPA and the conditional FDR approach at $N$ = 5000, $m$ = 1000.}\label{cond_AUC}
\end{figure}

As a comparison, we also tried the ``conditional FDR'' approach proposed by Andreassen et al. \cite{andreassen2013improved}  to prioritize the SNPs in our simulations. The comparison between this approach and GPA at $N = 5000$ and $m = 1000$ is shown in Figure \ref{cond_AUC}. GPA significantly outperformed the conditional FDR approach in terms of the SNP prioritization accuracy. More importantly, when no pleiotropy exists, the conditional FDR approach showed worse accuracy than single-GWAS analysis using the standard FDR approach,
whereas GPA achieved comparable accuracy with single-GWAS analysis in this scenario. This suggests that GPA was able take advantage of pleiotropy while it sacrifices much less statistical power than the conditional FDR approach when pleiotropy does not exist.

\begin{figure}
  \centering
  \includegraphics[width=.9\linewidth]{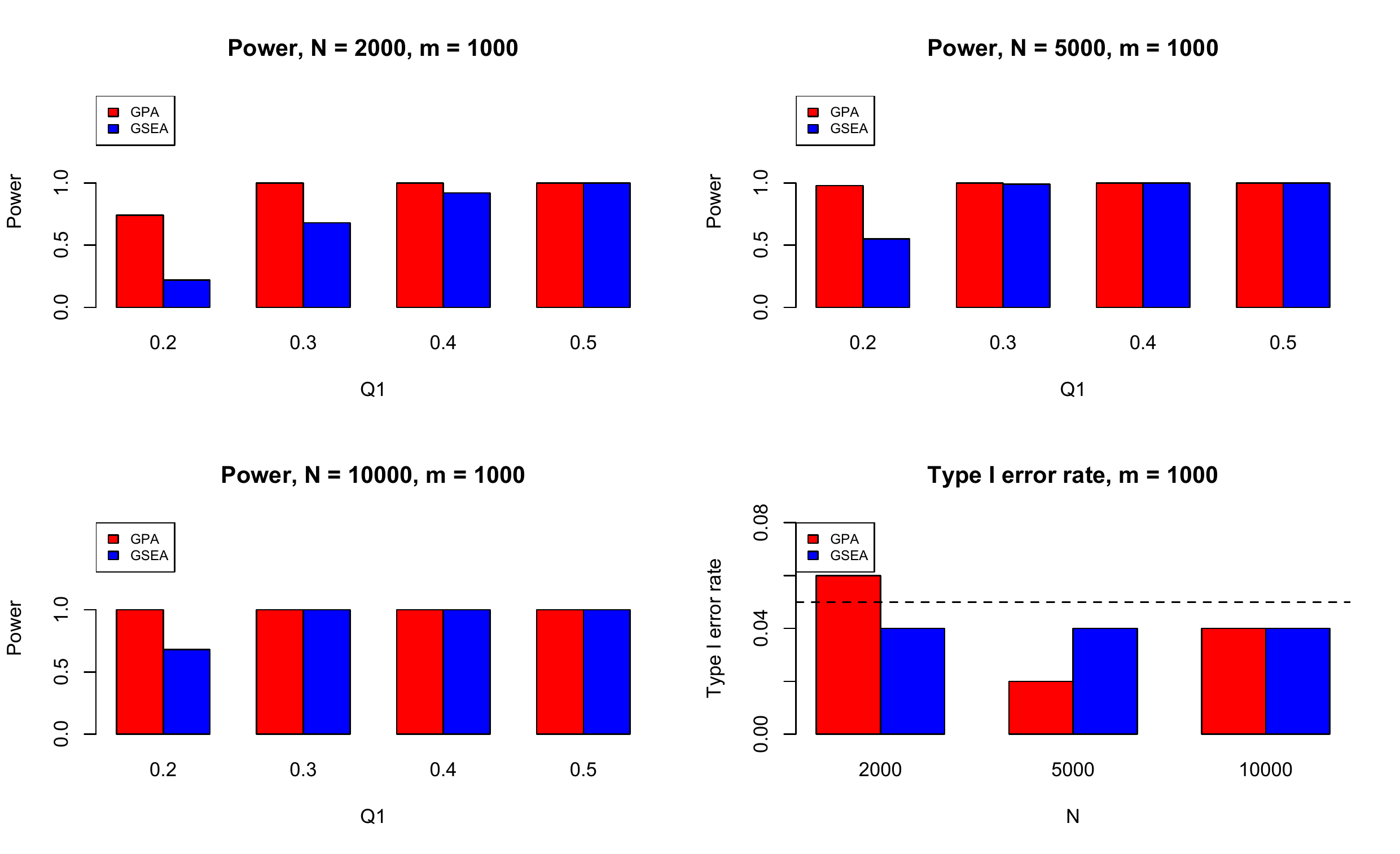}
  \caption{The comparison between GPA and GSEA at $m$ = 1000.}\label{compare_GSEA}
\end{figure}

Next, we evaluated the power and type I error of GPA for hypothesis testing on the significance of annotation enrichment in causal SNPs. Another popular method that does a similar job is the Gene Set Enrichment Analysis (GSEA) \cite{subramanian2005gene}. 
Although GSEA typically is used for analysis of gene expression data, its input can be a list of $p$-values obtained from any type of sources. Therefore we implemented the GSEA method to test the enrichment the $p$-values of a set of SNPs being annotated and compared it with GPA. We followed the previous simulation scheme and simulated one GWAS data set with $M = 20000$, $N$ varying from 2000 to 10000, $m$ varying from 500 to 2000. Here $q_0$ was fixed at 0.1 and $q_1$ was varied from 0.1 to 0.5. We performed tests at the significance level of 0.05. Type I error rate was evaluated at $q_1 = 0.1$ and power was evaluated in other cases. The result for $n = 1000$ was shown in Figure \ref{compare_GSEA}. We observed that in general,
GPA provided much higher power than GSEA while both methods appropriately controlled the type I error rate.

\begin{figure}
  \centering
  \includegraphics[width=.9\linewidth]{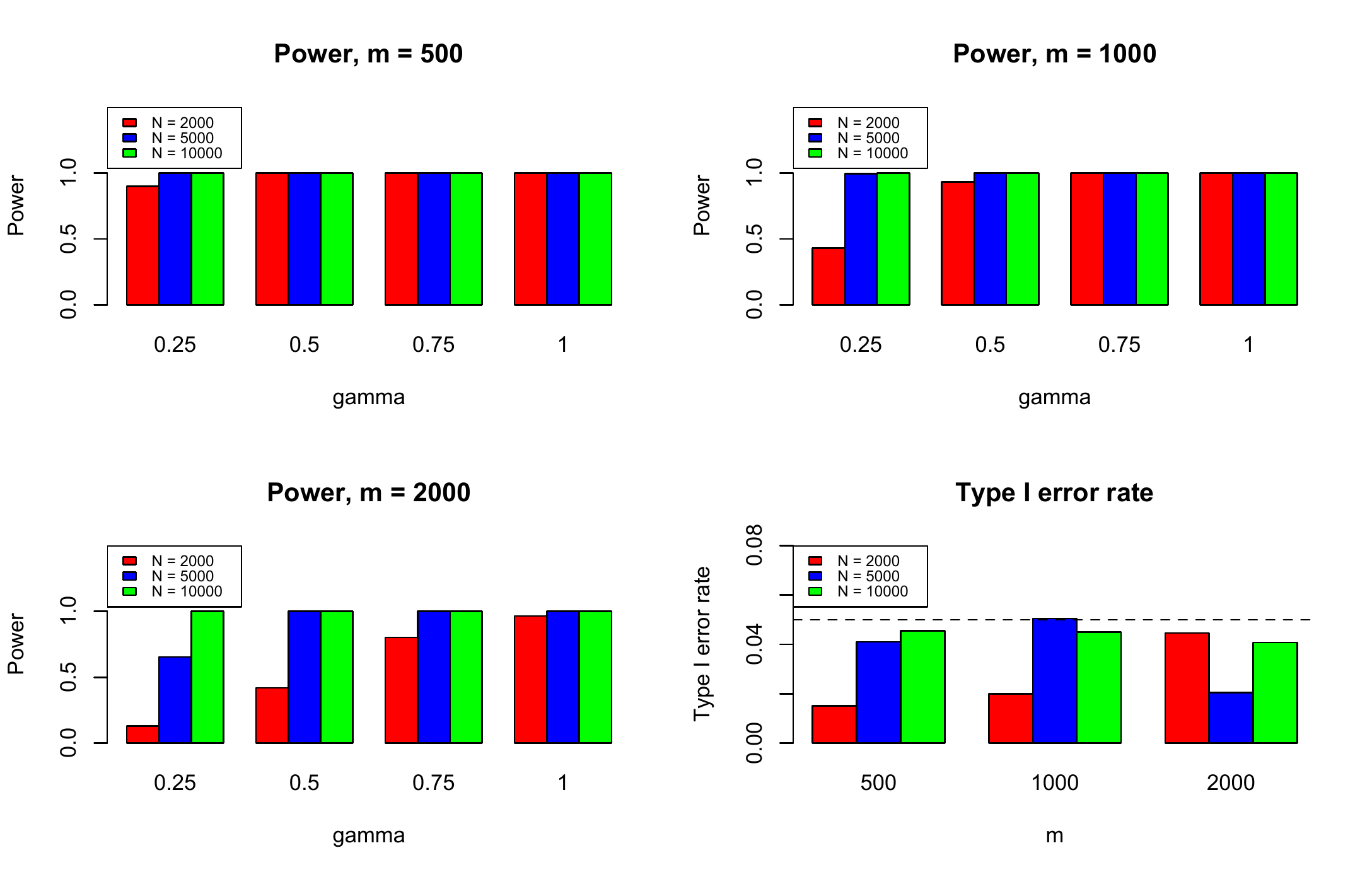}
  \caption{The power and type I error rate of the pleiotropy test.}\label{test_pleio}
\end{figure}

Lastly, we evaluated the power and type I error rate of GPA for the test of pleiotropy in our simulations. The simulation parameters were the same as in the previous simulations. Power was evaluated at $\gamma = 0.25$, $0.5$, $0.75$ and $1$. The type I error rate was evaluated at $\gamma = m/M$. As shown in Figure \ref{test_pleio}, the power increases as $m$ decreases and as $N$ and $\gamma$ increases, whereas the type I error rate was appropriately controlled in all cases.

\subsection{Real data analysis}

\subsubsection{GWAS of five psychiatric disorders}

We applied our GPA model to analyze the five psychiatric disorders \cite{pgc2013genetic,pgp2013improved}: attention deficit-hyperactivity disorder (ADHD), autism spectrum disorder (ASD), bipolar disorder (BPD), major depressive disorder (MDD) and schizophrenia (SCZ). Detailed information about these data sets is provided in \cite{pgc2013genetic,pgp2013improved}. The summary statistics of the five psychiatric disorders were downloaded from the section of cross-disorder at the website of Psychiatric Genomics Consortium (PGC) \url{https://pgc.unc.edu/Sharing.php}. The $p$-values were available for 1,230,535 SNPs in ADHD, 1,245,864 SNPs in ASD, 1,233,533 SNPs in BPD, 1,232,794 SNPs in MDD, and 1,237,959 SNPs in SCZ, respectively.
We took the intersection of those SNPs and obtained their $p$-values. Finally, we had a $p$-value matrix $\mathbf{P}\in \mathbb{R}^{1,219,805\times 5}$ for the five psychiatric disorders.

First, we performed single-GWAS data analysis using genes preferentially expressed in the central nervous system (CNS) \cite{lee2012estimating,raychaudhuri2010accurately} as the annotation data. Specifically, we generated the annotation vector $\bfA\in\mathbb{R}^{1,219,805}$ as follows: The entries in $\bfA$ corresponding to SNPs exactly in the genes from the CNS set or within 50-kb boundaries of those genes were set to be 1. Consequently, 21.9\% of the genome-wide SNP markers were annotated. The analysis results of these five psychiatric disorders were given in Table \ref{single-gwas-annotation}. The estimated fold enrichment $\hat{q}_1/\hat{q}_0$ of the CNS set was 1.749 (s.e. 0.447), 1.261 (s.e. 0.055), 1.467 (s.e. 0.033), 1.177 (s.e. 0.058) and 1.391 (s.e. 0.022)
 for ADHD, ASD, BPD, MDD and SCZ, respectively. PGC also evaluated enrichment of the CNS gene set by variance component estimation using linear mixed models (LMM) \cite{pgc2013genetic}, suggesting about 1.6\% and 1.5\% fold enrichment in BPD and SCZ, respectively.
 The minor difference of fold enrichment estimation between GPA and LMM could be understandable: First, GPA only used summary statistics while LMM used both phenotype and genotype data. Second, the mathematical definition of fold enrichment is different between GPA and LMM: GPA used the ratio between $\hat{q}_1$ and $\hat{q}_0$, while LMM used the ratio between the proportion of the variance explained by SNPs in the CNS set and the proportion of the CNS set in entire genome. Furthermore, we evaluated the significance of enrichment of the CNS set by hypothesis testing. As given in Table \ref{single-gwas-annotation}, enrichment of the CNS gene set was strong in BPD and SCZ, moderate in ASD and MDD, and nonsignificant in ADHD.

\begin{table}\scriptsize
  \centering
  \begin{tabular}{cccccccc}
    \hline
     & $\hat{\alpha}$ & $\hat{\pi}_0$ & $\hat{\pi}_1$ &$\hat{q}_0$ & $\hat{q}_1$ & $\hat{q}_1/\hat{q}_0$&$p$-value\\
    \hline
   ADHD  & 0.694 (0.103) & 0.991 (0.006)  & 0.009 (0.006) & 0.218 (0.001) & 0.381 (0.055) &  1.749 (0.447) & 0.083\\
   ASD  & 0.710 (0.014)  & 0.909 (0.007) & 0.091 (0.007) & 0.214 (0.001) & 0.270 (0.009) & 1.261 (0.055) & 8.408e-07\\
   BPD  & 0.697 (0.007) & 0.821 (0.007) & 0.179 (0.007) & 0.202 (0.001)  & 0.297 (0.004) & 1.467 (0.033) & 1.439e-48\\
   MDD  & 0.837 (0.019)  & 0.807 (0.027) & 0.193 (0.027) & 0.212 (0.003) & 0.249 (0.008) & 1.177 (0.058) & 0.005\\
   SCZ  & 0.596 (0.004)  & 0.804 (0.004) & 0.196 (0.004) & 0.203 (0.001) & 0.283 (0.003) & 1.391 (0.022) & 7.742e-79\\
    \hline
  \end{tabular}
  \caption{Single-GWAS analysis of five psychiatric disorders using the CNS gene set as the annotation data. Here $\hat{\alpha}$ is the estimate of $\alpha$ parameter of Beta distribution (\ref{twogroup2}), $\hat{\pi}_0$ and $\hat{\pi}_1$ are the estimated proportion of null-SNPs and non-null-SNPs defined in (\ref{twogroup2}), $\hat{q}_{0}$ and $\hat{q}_{1}$ are the estimated proportion of null and non-null SNPs in the CNS gene set. Enrichment fold $\hat{q}_1/\hat{q}_0$ and $p$-value given by hypothesis testing of enrichment in annotation data are provided in last two columns. The values in the brackets are standard errors of the estimates. }\label{single-gwas-annotation}
\end{table}

Next, we applied GPA to study pairwise pleiotropy of these five psychiatric disorders without using annotation data. As shown in Table \ref{tab_pleiotrop}, our analysis result suggested the pleiotropy effect was strong between BPD and SCZ ($p$-value is essentially zero),
MDD and SCZ ($p$-value $2.034\times 10^{-103}$), BDP and MDD ($p$-value $3.017\times 10^{-29}$), ASD and SCZ ($p$-value $5.749\times 10^{-25}$),
ASD-BPD ($p$-value $8.074\times 10^{-17}$),
moderate between ADHD and BPD ($p$-value $4.670\times 10^{-8}$), ADHD and SCZ ($6.837\times 10^{-5}$), and non-significant for other pairs. Our analysis result agrees with the most of pleiotropy results reported in \cite{pgc2013genetic} and the disagreement mainly came
from the joint analysis between ADHD and other disorders. The pleiotropy between ADHD and MDD was reported to be moderate, while GPA did not detect this moderate effect. From single GWAS analysis of ADHD, given in Table \ref{single-gwas-annotation}, the estimated parameters ($\hat{\pi}_1$=0.009 (s.e. 0.006), $\hat{\alpha}$=0.694) indicates that its GWAS signal was very weak. For MDD, the estimated parameter $\hat{\alpha}$=0.837 also indicates the weak marginal signal of MDD.
Consequently, the marginal GWAS signals of ADHD and MDD were too weak to allow GPA to detect the pleiotropy effect between them. Since the data analysis performed in \cite{pgc2013genetic} used genotype data, the bivariate linear mixed model could still have enough power to detect the moderate genetical correlation between ADHD and MDD.

\begin{table}\scriptsize
  \centering
\begin{tabular}{ccccccc}
  \hline
   & $\hat{\pi}_{00}$ & $\hat{\pi}_{10}$ & $\hat{\pi}_{01}$ & $\hat{\pi}_{11}$ & LRT & $p$-value \\
  \hline
  ADHD-ASD & 0.900 (0.009) & 0.007 (0.006) & 0.093 (0.009) & 0.001 (0.004) & 0.913 & 0.339\\
  ADHD-BPD & 0.822 (0.008) & 0.001 (0.005) & 0.164 (0.009) & 0.013 (0.007) & 29.849 & 4.670e-08\\
  ADHD-MDD & 0.776 (0.036) & 0.006 (0.010) & 0.217 (0.036) & 0.001 (0.010) & -0.005 & 1\\
  ADHD-SCZ & 0.804 (0.005) & 0.001 (0.004) & 0.183 (0.008) & 0.012 (0.007) & 15.855 & 6.837e-05\\
  ASD-BPD & 0.791 (0.008) & 0.027 (0.007) & 0.115 (0.009) & 0.067 (0.008) & 69.391 & 8.074e-17\\
  ASD-MDD & 0.727 (0.033) & 0.049 (0.016) & 0.180 (0.033) & 0.044 (0.016) & 2.717 & 0.099\\
  ASD-SCZ & 0.771 (0.006) & 0.035 (0.006) & 0.131 (0.007) & 0.064 (0.006) & 106.493 & 5.749e-25\\
  BPD-MDD & 0.793 (0.014) & 0.011 (0.026) & 0.030 (0.015) & 0.166 (0.027) & 126.037 & 3.017e-29\\
  BPD-SCZ & 0.821 (0.004) & 0.001 (0.005) & 0.013 (0.006) & 0.165 (0.007) & 1851.727 & 0\\
  MDD-SCZ & 0.809 (0.009) & 0.001 (0.010) & 0.001 (0.025) & 0.189 (0.025) & 466.312 & 2.034e-103\\
  \hline
\end{tabular}
  \caption{Pleiotropy estimated among five psychiatric disorders. The values in the brackets are standard errors of the estimates. The last two columns provide the LRT statistics and $p$-values of hypothesis testing (\ref{hypotheses2}).}\label{tab_pleiotrop}
\end{table}


We further applied GPA to study all pairs of disorders using the CNS gene set as the annotation data. The estimated $\hat{q}_l$ ($l\in\{00,10,01,11\}$) are given in Table \ref{two-gwas-annotation} and $\hat{\pi}_l$ remained almost the same as without annotation data. The $p$-values of hypothesis test (\ref{joint-ann-hypo}) are also provided in the last column of Table \ref{two-gwas-annotation}. The p-value should be interpreted with caution: as shown in Table \ref{single-gwas-annotation}, the CNS gene set is enriched in all these disorders except ADHD. Hence, the significant $p$-values listed in Table \ref{two-gwas-annotation} may be simply due to the combinatorial effects. On the other hand, the ratio between $\hat{q}_{11}$ and $\hat{q}_{00}$ could be more interesting. Take BPD-SCZ as an example. The ratio between $\hat{q}_{11}$ and $\hat{q}_{00}$ is 1.503 (s.e. 0.025), which suggests that enrichment of the CNS set for the BDP-SCZ shared risk variants was even stronger than that for BPD-only (1.467 (s.e. 0.033)) or SCZ-only (1.391 (s.e. 0.022)).

\begin{table}\scriptsize
  \centering
  \begin{tabular}{cccccc}
    \hline
     & $\hat{q}_{00}$ & $\hat{q}_{10}$ & $\hat{q}_{01}$ & $\hat{q}_{11}$ & $p$-value\\
    \hline
    ADHD-ASD & 0.212 (0.002) & 0.425 (0.146) & 0.272 (0.013) & 0.022 (1.879) & 4.205e-06\\
    ADHD-BPD & 0.202 (0.003) & 0.975 (0.261) & 0.304 (0.010) & 0.216 (0.143) & 2.965e-47\\
    ADHD-MDD & 0.209 (0.004) & 0.490 (0.411) & 0.251 (0.014) & 0.001 (2.357) & 0.005 \\
    ADHD-SCZ & 0.204 (0.002) & 0.001 (2.349) & 0.261 (0.015) & 0.511 (0.100) & 7.132e-79 \\
    ASD-BPD & 0.204 (0.003) & 0.164 (0.073) & 0.285 (0.015) & 0.318 (0.022) & 2.058e-51 \\
    ASD-MDD & 0.193 (0.006) & 0.470 (0.069) & 0.309 (0.018) & 0.002 (0.171) &  3.251e-09\\
    ASD-SCZ & 0.199 (0.002) & 0.295 (0.028) & 0.296 (0.008) & 0.255 (0.018) & 2.078e-81\\
    BPD-MDD & 0.195 (0.004) & 0.568 (0.025) & 0.367 (0.066) & 0.222 (0.023) & 7.314e-48\\
    BPD-SCZ & 0.206 (0.001) & 0.001 (6.907) & 0.026 (0.720) & 0.309 (0.006) & 2.860e-130\\
    MDD-SCZ & 0.204 (0.002) & 0.001 (16.882) & 0.732 (1.019) & 0.260 (0.011) & 7.324e-78\\
    \hline
  \end{tabular}
    \caption{GPA results for all pairs of the five psychiatric disorders with the CNS set as annotation.
    The estimated $\hat{\pi}_l$ almost remain the same as Table \ref{tab_pleiotrop} and $\hat{q}_l$ are shown in the table.
    The values in the brackets are standard errors of the estimates.
    The $p$-values of hypothesis testing (\ref{joint-ann-hypo}) are provided in the last column.}\label{two-gwas-annotation}
\end{table}

\begin{figure}
  \centering
\includegraphics[width=.45\linewidth]{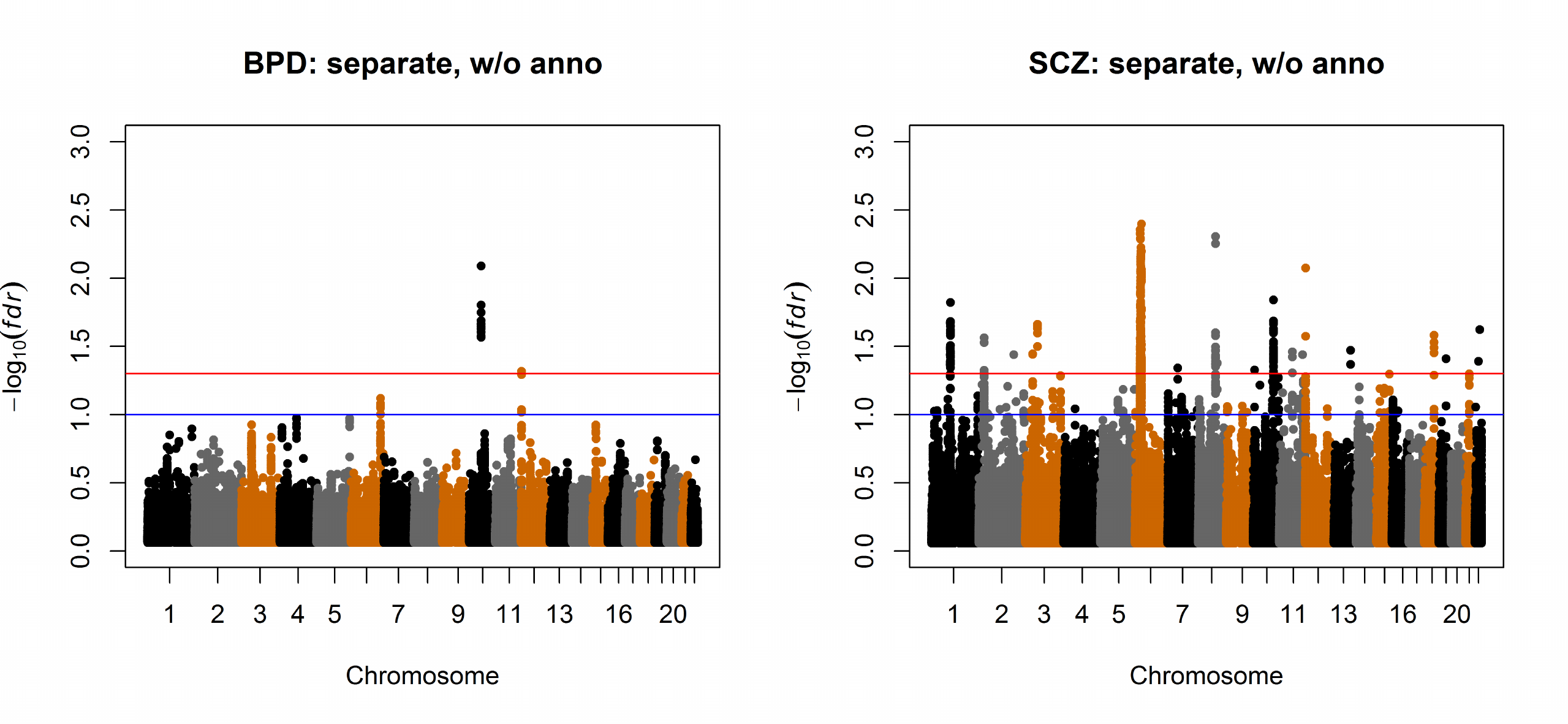} \hspace{2em}
\includegraphics[width=.45\linewidth]{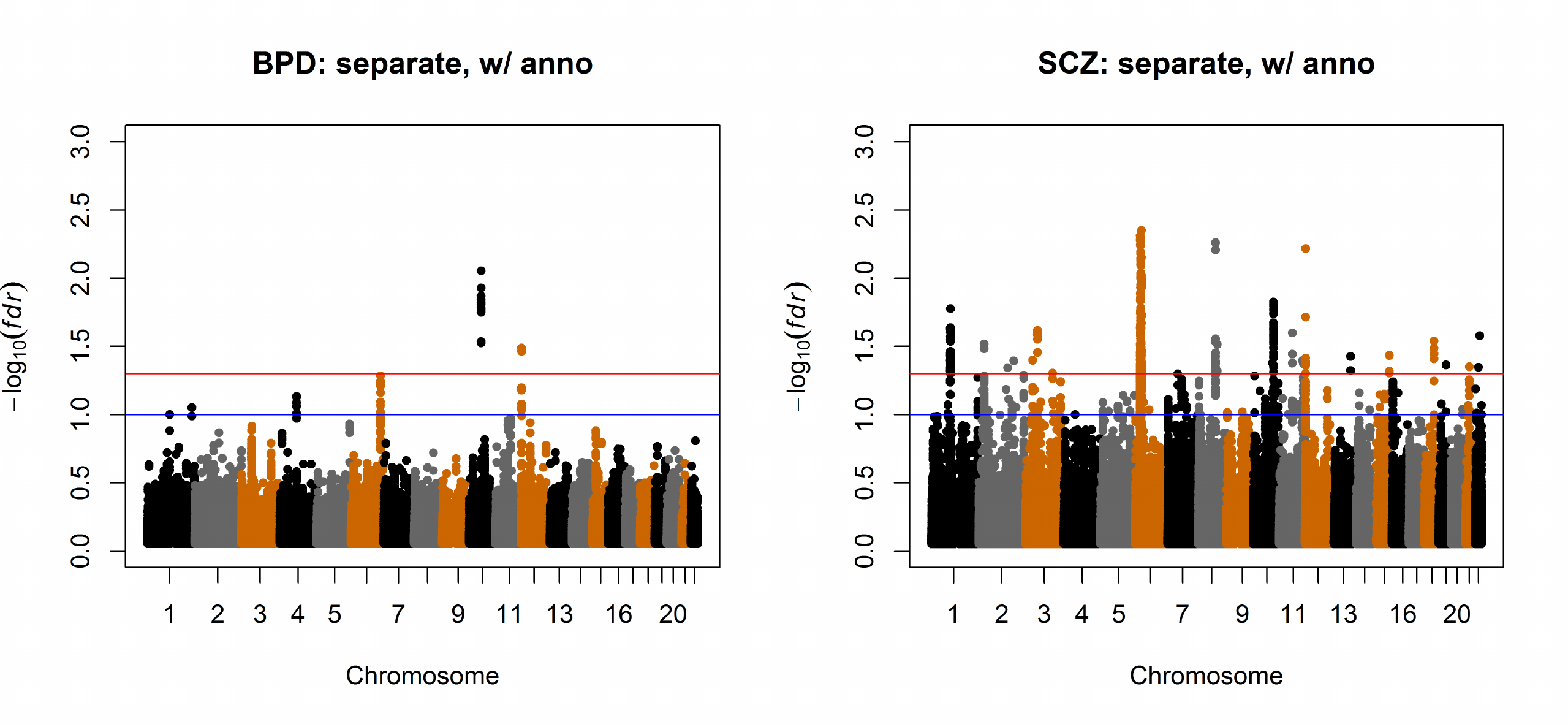} \\
\includegraphics[width=.45\linewidth]{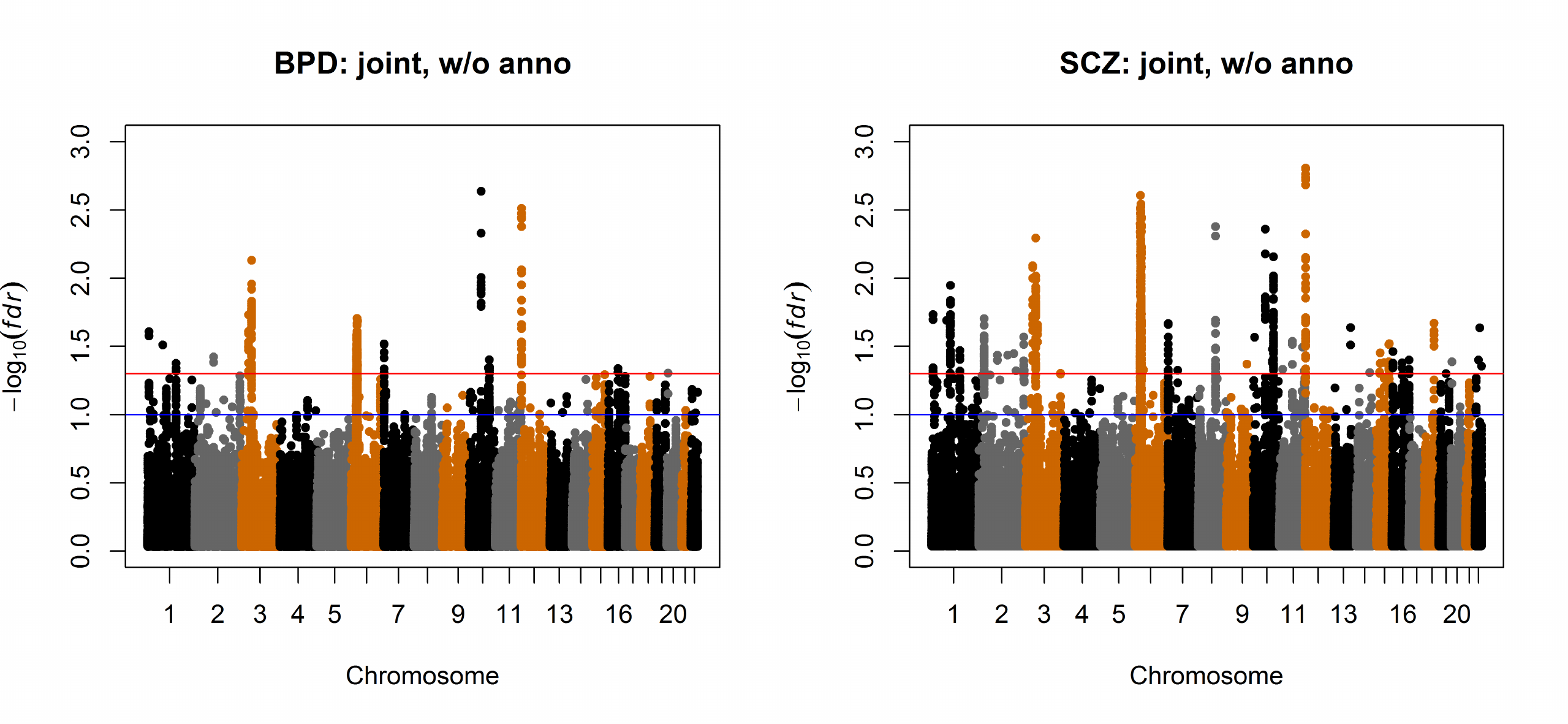} \hspace{2em}
\includegraphics[width=.45\linewidth]{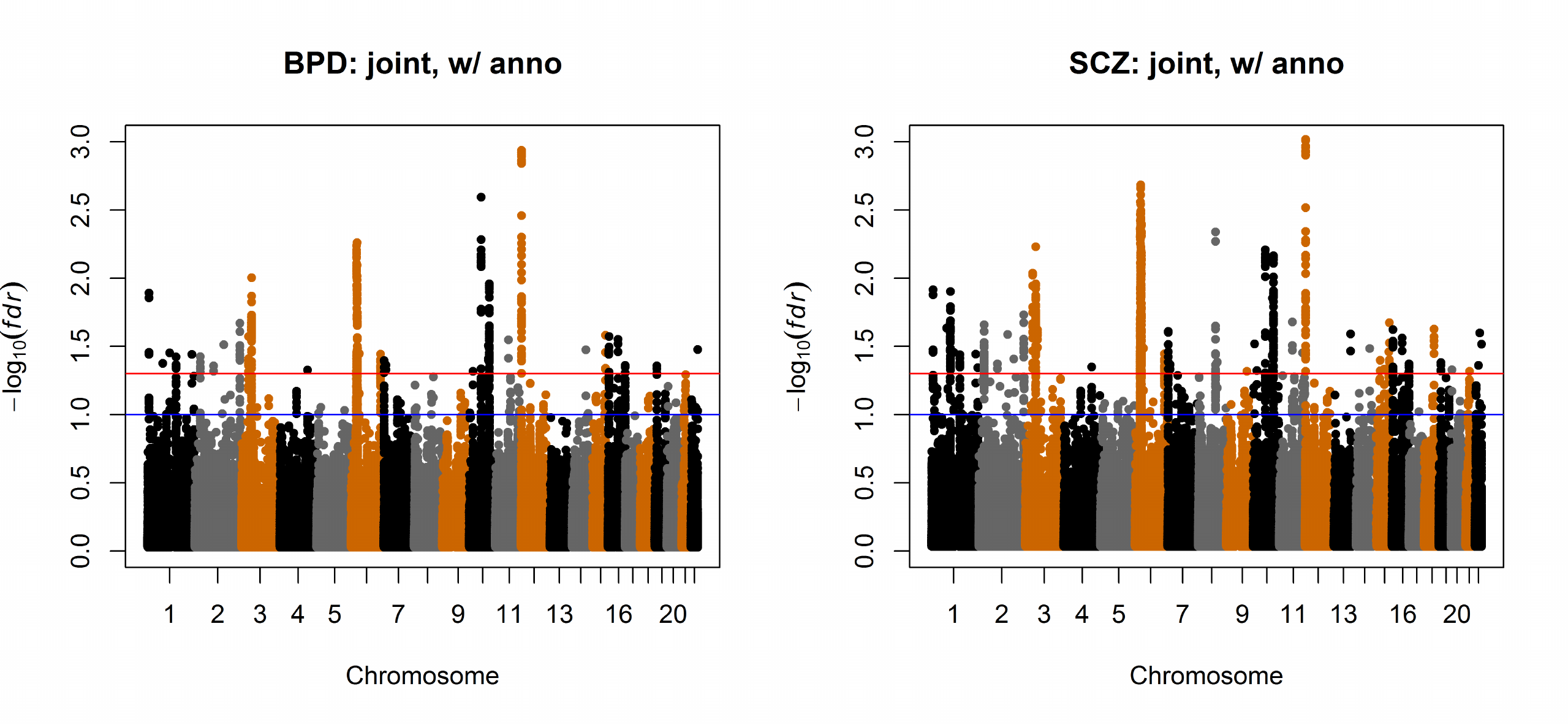} 
  \caption{Manhattan plots of BPD and SCZ. Top left panel: separate analysis without annotation.
  Top right panel: separate analysis with CNS
annotation. Bottom left panel: joint analysis without
annotation. Bottom right panel: joint analysis with CNS
annotation. The red and blue lines indicate $fdr$ = 0.05 and 0.1, respectively.}\label{BPD-SCZ}
\end{figure}

We also compared the results given by four different analysis approaches: single-GWAS analysis with or without annotation, two-GWAS joint analysis with or without annotation data. The manhattan plots are shown in Figure \ref{BPD-SCZ}. For single-GWAS analysis without annotation, GPA identified 13 SNPs and 391 SNPs with  $fdr<0.05$ for BPD and SCZ, respectively.  By using the CNS set as annotation, GPA was able to identify 14 and 409 SNPs for BPD and SCZ, respectively.
 For joint analysis without annotation, the number of identified SNPs dramatically increased to 383 and 821 for BPD and SCZ, respectively.  By using the CNS set as annotation, the number of identified SNPs further increased to 385 and 837 for BPD and SCZ, respectively.
We investigated the result of BPD more in detail to evaluate the power of GPA in identification of functionally important SNPs. For single-GWAS analysis of BPD, GPA was able to identify SNPs locating the  \emph{ANK3} gene. By using annotation data,
the  \emph{CACNA1C} gene which encodes an alpha-1 subunit of a voltage-dependent calcium channel was identified by GPA. After incorporating pleiotropy information between SCZ and BPD,
more functionally relevant genes, such as \emph{PBRM1}, \emph{C6orf136}, \emph{DPCR1}, \emph{SYNE1}, can be further identified by GPA. For instance, \emph{SYNE1} encodes the synaptic nuclear envelope protein 1, and provides instructions for making a protein called Syne-1 that is found in many tissues and especially critical in the brain. The Syne-1 protein is active (expressed) in Purkinje cells, which are located in the cerebellum and are involved in chemical signaling between nerve cells (neurons).
Mutations in the \emph{SYNE1} gene have been found to cause autosomal recessive cerebellar ataxia type 1 (ARCA1) and \emph{SYNE1} has recently been implicated as a susceptibility gene for BPD  in a large collaborative GWAS study \cite{sklar2011large}. Clearly, these results indicate that the statistical power to identify associated SNPs increased a lot by making use of pleiotropy and functional annotation (in this real data example, pleiotropy played a more important role than functional annotation).

We also tried to apply GPA for joint analysis of BDP, SCZ and MDD to further explore their joint genetic architecture.
In this scenario, GPA used eight-group to model all $2^3$ states of a SNP, which is a nature extension of four-group model (\ref{fourgroup}).
However, we found that the estimated parameters obtained from this model
were not reliable because of their large standard errors. Also, the interpretation of the estimated parameters became more complicated. Therefore,
we mainly focused on two-GWAS analysis in this paper.



Regarding to computational time, the GPA algorithm takes less than 20 minutes to analyze typical GWAS data sets.
The speed of convergence depends on the strength of the GWAS signals. For example, it took about 7 mins and 3 mins to
analyze ADHD and SCZ, repectively, as SCZ has a stronger GWAS signal than ADHD. For joint analysis of BPD and SCZ, it took about 20 mins.
All timings were carried out on a desktop with 3.0 GHz CPU with 16G memory.

\subsection{Bladder cancer GWAS and ENCODE annotation data}

In molecular biology, DNase I hypersensitive sites (DHSs) are regions where DNA degradation by enzymes like DNase I raise more frequently. As a result, DHSs can mark active transcription regions across genome and these patterns are known to be tissue or cell specific. The ENCODE project analyzed the DHSs in 125 human cell lines with the intention of cataloging human regulatory DNA \cite{thurman2012accessible}. In this section, we applied GPA to assess how bladder cancer \cite{rothman2010multi} risk associated SNPs are enriched in DHSs region across these 125 human cell lines.

We downloaded genotype data for bladder cancer from dbGaP (NCI Cancer Genetic Markers of Susceptibility (CGEMS) project; accession number phs000346.v1.p1). We used the samples genotyped from both Illumina 1M chip and 610K chip for our analysis. For quality control, we removed SNPs with a missing rate $>$ 0.01. We checked Hardy-Weinberg Equilibrium and excluded SNPs with $p$-value $<$ 0.001. SNPs with minor allele frequency (MAF) $<5\%$ were also removed. Finally, 490,614 SNPs from 3,631 cases and 3,356 controls of European descent were used in the analysis. Then, we analyzed this bladder cancer data with logistic regression by assuming additive model and obtained $p$-values. We also downloaded the uniform peak files for DHS in 125 cell lines from the ENCODE database (\url{http://genome.ucsc.edu/cgi-bin/hgFileUi?db=hg19&g=wgEncodeAwgDnaseUniform}). Note that the DHS for these 125 cell lines were identified with a uniform analysis workflow by the ENCODE Consortium and this facilitates fair and unbiased comparison among cell lines as annotation for our GPA model.

\begin{figure}
  \centering
  \includegraphics[width=.48\linewidth]{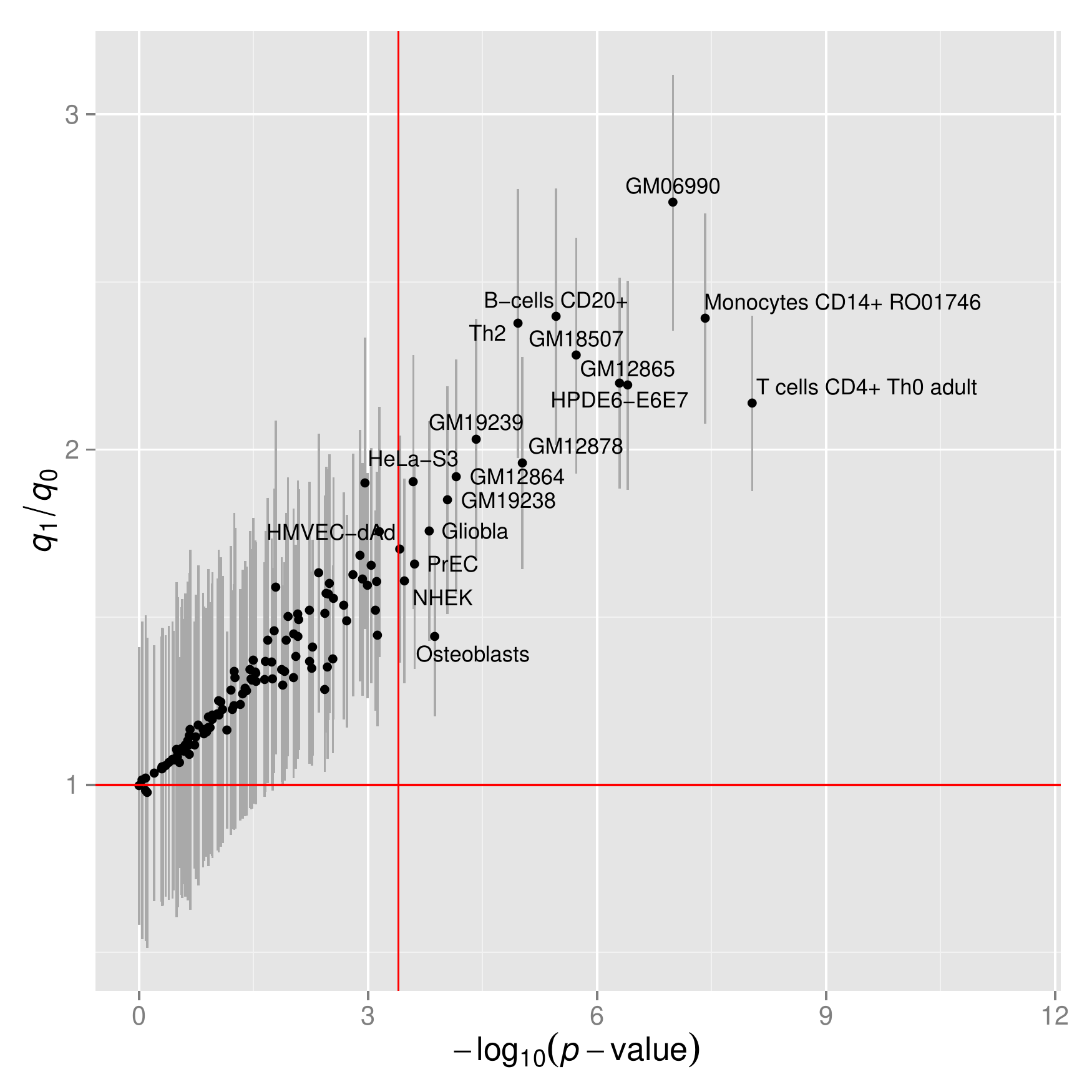}
  \includegraphics[width=.48\linewidth]{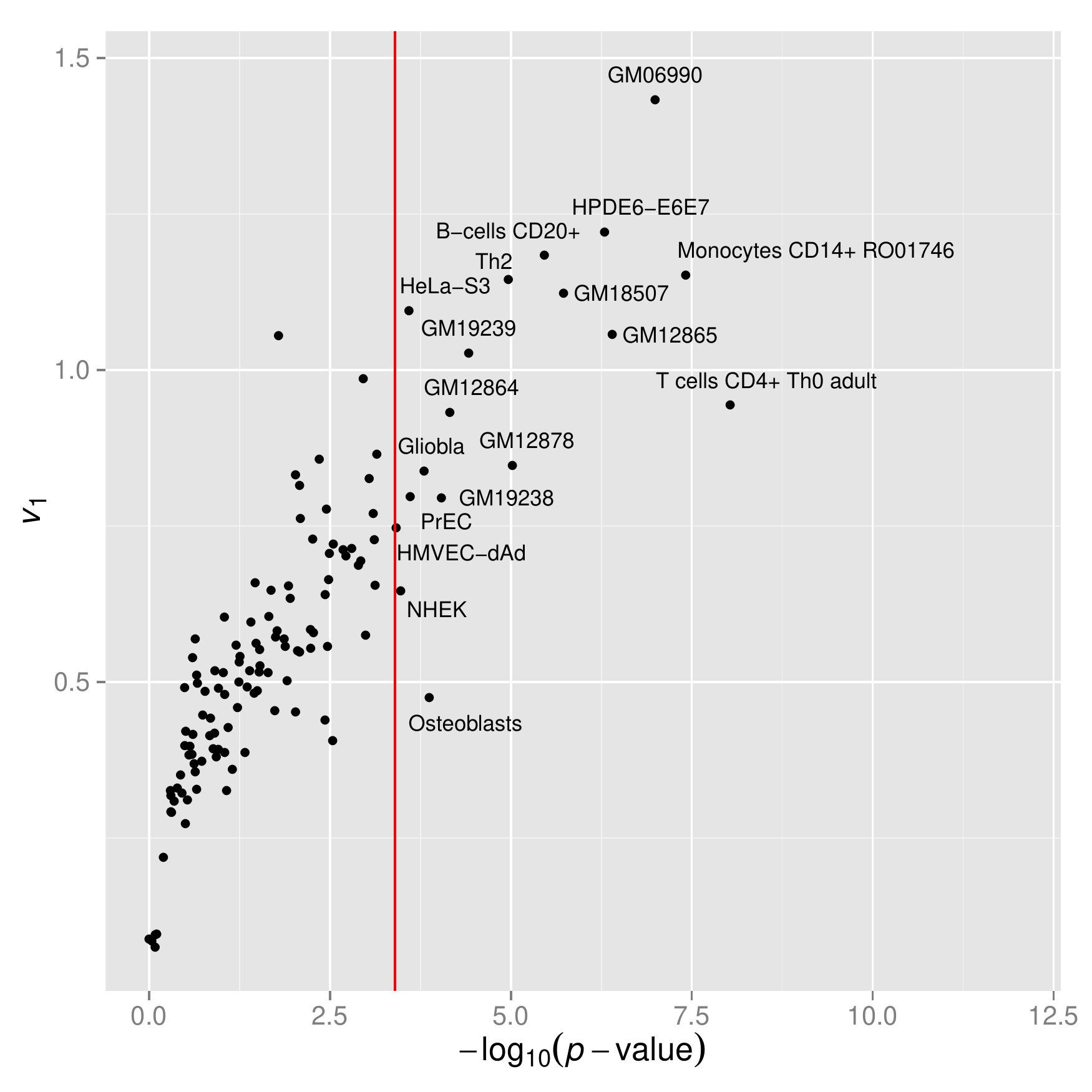}
  \caption{Enrichment of the DNase I hypersenstivity site annotation data from 125 cell lines for bladder cancer. Left panel: $-\log(p-\mbox{value})$ of hypothesis testing (\ref{hypotheses0}) vs. fold enrichment $\hat{q}_{1}/\hat{q}_{0}$. The vertical red line corresponds to the significance level ($\alpha$ = 0.05) after Bonferroni correction. The horizontal red line corresponds to  ratio=1. Right panel: The normalized variance component $v_1$ (\ref{v1}) given by LMM v.s. $-\log(p-\mbox{value})$  given by GPA.}\label{encode}
\end{figure}

We applied GPA to analyze the bladder cancer GWAS data set with one annotation data at a time, and performed hypothesis testing to assess the significance of enrichment.
The result is shown in the left panel of Figure \ref{encode}. Under significance level $\alpha = 0.05$ after Bonferroni correction,
annotations from 19 cell lines were significantly enriched for bladder cancer risk associated SNPs.
Most of these cell lines were derived from lymphocytes from normal blood (e.g., T cells CD4+ Th0 adult, Monocytes CD14+ RO01746),
while some cell lines came from cancer patients (e.g., Gliobla and HeLa-S3).
The above results demonstrates that the functional roles of bladder cancer risk variants might be involved in disturbance
of immune system or carcinoma pathways. This also implies that GPA may be an effective way to explore functional role of
GWAS hits by testing enrichment on phenotype-related annotations or user-specified annotations.

 We also compared GPA with the LMM-based approach \cite{lee2011estimating,yang2011genome}. Specifically, we considered the following genome-partitioning linear mixed model:
\begin{align}\label{}
& \bfy = \bfX \bfbeta + \bfW_1 \bfu_1 + \bfW_2 \bfu_2 + \bfe, \nonumber\\
& \bfu_1 \sim \mathcal{N}( 0, \sigma_1^2 \bfI ), \bfu_2 \sim \mathcal{N}( 0, \sigma_2^2 \bfI ), \bfe \sim \mathcal{N}( 0, \sigma_e^2 \bfI ),
\end{align}
where $\bfX$ are covariates (the first five principal components from genotype data), and $\bfW_1$ and $\bfW_2$ are sets of SNPs overlapping DHS in each cell line and the remaining SNPs, respectively. We denote number of SNPs in $\bfW_1$ and $\bfW_2$ as $M_1$ and $M_2$, respectively. Median number of SNPs that overlap DHS in each cell line is about 60K and 90\% of cell lines have number of DHS ranging between 40K and 80K. In order to take into account such variation in number of DHS among cell lines, we defined a scaled version of the proportion of phenotype variance explained by SNPs overlapping DHS in each cell line as
\begin{align}\label{v1}
v_1 = \frac{ M }{ M_1 } \frac{ M_1 \sigma_1^2 }{ ( M_1 \sigma_1^2 + M_2 \sigma_2^2 + \sigma_e^2 ) },
\end{align}
where $\frac{ M_1 \sigma_1^2 }{ ( M_1 \sigma_1^2 + M_2 \sigma_2^2 + \sigma_e^2 ) }$ is the proportion of the explained variance and
 $\frac{ M }{ M_1 }$ is the scaling factor.
The right panel of Figure \ref{encode} shows that ($-log_{10}$)-transformed $p$-value of the GPA annotation enrichment test is well linearly related to $v_1$. This indicates that our GPA model captures enrichment of annotation almost as accurately as LMM even without the original genotype data, which implies wide applicability of our GPA model.

\section{Discussion and Conclusion}

\subsection{Relationship between GPA and other related methods}
In this subsection, we briefly discuss the relationship of GPA with other related methods, such as LMM, conditional FDR and GSEA.

LMM is an effective tool for exploring genetic architecture of complex traits/diseases and it has been implemented in a popular software
named GCTA \cite{yang2011gcta}. Compared with LMM, GPA has the following distinct features:
\begin{itemize}
  \item LMM explores the genetic architecture underlying of complex trait/disease by estimating the gross phenotypic variance
  that can be explained by whole genome or a certain subset of SNPs.
  In contrast, GPA provides more ``fine-grained'' understanding by giving estimates of the local fasle discovery rate of each SNP, the proportion of SNPs that
   are associated with the phenotype ($\{\pi_l\}$), the overall effect strength of the associated SNPs ($\alpha$), and enrichment of
   a particular functional annotation ($\{q_l\}$).
  \item Application of LMM requires the availability of genotype data, while GPA only needs the summary statistics ($p$-values) as its input.
  Typically, the genotype data may not be accessible as easily as the summary statistics. For example, when researchers want to
  implement the integrative analysis for their own GWAS data at hand with related GWAS studies,
   it is much easier for them to obtain the summary statistics than the whole data sets of related GWAS studies.
  In this sense, GPA may greatly simplify the procedure of integrative analysis.
\end{itemize}

To our best knowledge, the conditional FDR approach is the first approach that statistically addresses the issue of pleiotropy between two GWAS,
and GSEA is the most popular approach to evaluating the enrichment of gene sets. In fact, GPA provides a unified framework for systematically
integrating both sources of information: pleiotropy and annotation. Rigorous statistical inference of pleiotropic effects and annotation enrichment
has been established in this framework. As demonstrated in our simulation study (based on the widely accepted random-effects model in GWAS),
GPA has better performance of identifying disease-associated markers than the conditional FDR approach, and it shows a greater power of evaluating
annotation enrichment than GSEA as well.

\subsection{Conclusion}
A large number of GWAS have been carried out in the last few years. As the cumulation of GWAS data, it is time to investigate systematic
analysis of GWAS data sets to provide comprehensive understanding of the genetic architecture of complex traits/diseases, and provide new insights
for functional genome studies.
To achieve this goal, there have been fast growing interests in developing computational and statistical
approaches to exploring genomic data in the post-GWAS era. In this paper, we
present our approach, named GPA, which integrates information from multiple GWAS data sets and functional annotation data. Not only
has GPA a better statistical power than related methods, it also provides interpretable model parameters which deepens our understanding
of the genetic architecture of complex traits/diseases. We have successfully applied GPA to analyze GWAS data of five psychiatric disorders from PGC.
The real data results demonstrate that GPA is able to identify pleiotropic effects among psychiatric disorders and detect enrichment of the CNS gene set.
We have also applied GPA to analyze bladder cancer GWAS data with ENCODE data as annotation, where the significant enrichment of immune system and carcinoma pathways
has been observed. In addition, GPA shows similar results of enrichment analysis to LMM, even without the genotype data. This suggests
that GPA can serve as an effective tool for the integrative analysis of multiple GWAS data with functional annotation data, when genotype data are not available.

In this work, we mainly focused on studying the role of a single annotation vector. To deal with multiple annotation vectors, we assumed
the conditional independence among them (see GPA model (\ref{GPA_model})). As a result, the current version of GPA may become unreliable in the presence of
multiple highly correlated annotation vectors. Hence, when multiple sources of annotation data are available, we suggest incorporating them into GPA
one at a time, as demonstrated in our analysis of bladder cancer GWAS data with ENCODE annotation. We realize that simultaneous analysis of
multiple correlated annotation vectors may be more powerful. We will investigate
this issue in our future work.

\section*{Web source}

The R package of GPA is publicly available from GitHub http://dongjunchung.github.io/GPA/ and will be available in Bioconductor upon acceptance of the manuscript.

\section*{Appendix}

\subsection*{The Expectation-Maximization algorithm}

In order to be consistent with the main text, we present the EM algorithm for the case that we have two GWAS data but the generalization to more than two GWAS data is straightforward and the actual GPA algorithm is not limited to the number of GWAS data.
In the GPA model, we have the parameter vector of length $( 2^K + K + 2^K D )$ as
\begin{eqnarray*}
\Theta = ( \pi_{00}, \pi_{10}, \pi_{01}, \pi_{11}, \alpha_1, \alpha_2, [ q_{d00}, q_{d10}, q_{d01}, q_{d11} ]_{d \in \{1, \cdots, D\}} )
\end{eqnarray*}
and the complete likelihood as
\begin{eqnarray*}
L_c ( \Theta ) = \prod_{j=1}^M \prod_{l\in\{00,10,01,11\}} \left[ \pi_{l} \pr(P_{j1},P_{j2}|Z_{jl} = 1 ; \Theta)
\prod_{d=1}^D \pr({A}_{jd}|Z_{jl}=1 ; \Theta) \right]^{ Z_{jl} }.
\end{eqnarray*}
If annotation data is not incorporated, this complete likelihood is simplified to
\begin{eqnarray*}
L_c ( \Theta ) = \prod_{j=1}^M \prod_{l\in\{00,10,01,11\}} \left[ \pi_{l} \pr(P_{j1},P_{j2}|Z_{jl} = 1 ; \Theta) \right]^{ Z_{jl} }.
\end{eqnarray*}
Based on this complete likelihood, the E- and M-steps in $t$-th iteration of the EM algorithm are obtained as follows.\\

\noindent
\textbf{E-step:}\\

\noindent
For $l\in\{00,10,01,11\}$, posterior probabilities for association of $j$-th SNP are obtained as:
\begin{align*}
z_{jl}^{(t)} &= \pr( Z_{jl} = 1 | \mathbf{P}, \mathbf{A} ; \Theta^{(t)} )\\
&= \frac{ \pi_{l}^{(t)} \pr(P_{j1},P_{j2}|Z_{jl} = 1 ; \Theta^{(t)}) \prod_{d=1}^D \pr({A}_{jd}|Z_{jl}=1 ; \Theta^{(t)}) }{ \sum_{l \prime \in\{00,10,01,11\}} \pi_{l \prime}^{(t)} \pr(P_{j1},P_{j2}|Z_{jl \prime} = 1 ; \Theta^{(t)}) \prod_{d=1}^D \pr({A}_{jd}|Z_{jl \prime}=1 ; \Theta^{(t)}) }.
\end{align*}

\noindent
If annotation data is not incorporated, for $l\in\{00,10,01,11\}$, we have
\begin{align*}
z_{jl}^{(t)} = \pr( Z_{jl} = 1 | \mathbf{P} ; \Theta^{(t)} )
  = \frac{ \pi_{l}^{(t)} \pr(P_{j1},P_{j2}|Z_{jl} = 1 ; \Theta^{(t)}) }{ \sum_{l \prime \in\{00,10,01,11\}} \pi_{l \prime}^{(t)} \pr(P_{j1},P_{j2}|Z_{jl \prime} = 1; \Theta^{(t)}) }.
\end{align*}

\noindent
\textbf{M-step:}\\

\noindent
Parameters for proportion of SNPs in each association status category are estimated as:
\begin{eqnarray*}
\pi_{00}^{(t+1)} = \frac{1}{M} \sum_{j=1}^M z_{j00}^{(t)},
\pi_{10}^{(t+1)} = \frac{1}{M} \sum_{j=1}^M z_{j10}^{(t)},
\pi_{01}^{(t+1)} = \frac{1}{M} \sum_{j=1}^M z_{j01}^{(t)},
\pi_{11}^{(t+1)} = \frac{1}{M} \sum_{j=1}^M z_{j11}^{(t)}.
\end{eqnarray*}

\noindent
Parameters for enrichment of $d$-th annotation data for association of SNPs are estimated as:
\begin{eqnarray*}
q_{d00}^{(t+1)} = \frac{ \sum_{j=1}^M z_{j00}^{(t)} A_{jd} }{ \sum_{j=1}^M z_{j00}^{(t)} },
q_{d10}^{(t+1)} = \frac{ \sum_{j=1}^M z_{j10}^{(t)} A_{jd} }{ \sum_{j=1}^M z_{j10}^{(t)} },
q_{d01}^{(t+1)} = \frac{ \sum_{j=1}^M z_{j01}^{(t)} A_{jd} }{ \sum_{j=1}^M z_{j01}^{(t)} },
q_{d11}^{(t+1)} = \frac{ \sum_{j=1}^M z_{j11}^{(t)} A_{jd} }{ \sum_{j=1}^M z_{j11}^{(t)} }.
\end{eqnarray*}

\noindent
Parameters for signal strength of GWAS data are estimated as:
\begin{eqnarray*}
\alpha_1^{(t+1)} = \frac{ \sum_{j=1}^M ( z_{j10}^{(t)} + z_{j11}^{(t)} ) }{ \sum_{j=1}^M ( z_{j10}^{(t)} + z_{j11}^{(t)} ) ( - log P_{j1} ) },
\alpha_2^{(t+1)} = \frac{ \sum_{j=1}^M ( z_{j01}^{(t)} + z_{j11}^{(t)} ) }{ \sum_{j=1}^M ( z_{j01}^{(t)} + z_{j11}^{(t)} ) ( - log P_{j2} ) }.
\end{eqnarray*}

\noindent
The M step remains the same when annotation data is not incorporated, except that we do not need to calculate $q_{d00}^{(t+1)}$, $q_{d10}^{(t+1)}$, $q_{d01}^{(t+1)}$, and $q_{d11}^{(t+1)}$.

\subsection*{Estimation of false discovery rate}

For analysis of single GWAS without annotation data, the local false discovery rate can be calculated as
\begin{equation}\label{fdr1}
  \mbox{fdr}(P_j)=\hat \pr(Z_{j0}=1|P_j) = \frac{\hat{\pi}_0 \pr(P_j|Z_{j0} = 1 ; \hat \Theta)}{\hat{\pi}_0 \pr(P_j|Z_{j0} = 1 ; \hat \Theta) + \hat{\pi}_1 \pr(P_j|Z_{j1} = 1 ; \hat \Theta)},
\end{equation}
where $\hat{\pi}_0$, $\hat{\pi}_1$, and $\hat \Theta$ are estimated from the EM algorithm.

For joint analysis of two GWAS data sets, we are interested in the local false discovery rate of the $j$-th SNP, if it is claimed to be associated with the first phenotype and the second one, i.e.,
\begin{align}\label{}
  &\mbox{fdr}_1(P_{j1},P_{j2})=\hat \pr(Z_{j00}+Z_{j01}=1|P_{j1},P_{j2})=\frac{\pr(P_{j1},P_{j2},Z_{j00}+Z_{j01}=1 ; \hat \Theta)}{\pr(P_{j1},P_{j2} ; \hat \Theta)},\nonumber\\
  &\mbox{fdr}_2(P_{j1},P_{j2})=\hat \pr(Z_{j00}+Z_{j10}=1|P_{j1},P_{j2})=\frac{\pr(P_{j1},P_{j2},Z_{j00}+Z_{j10}=1 ; \hat \Theta)}{\pr(P_{j1},P_{j2} ; \hat \Theta)},
\end{align}
where
\begin{align}\label{}
  \pr(P_{j1},P_{j2} ; \hat \Theta) &= \sum_{l\in\{00,10,01,11\}} \hat{\pi}_l\pr(P_{j1},P_{j2}|Z_{jl} ; \hat \Theta),\nonumber\\
  \pr(P_{j1},P_{j2},Z_{j00}+Z_{j01}=1 ; \hat \Theta) &= \sum_{l\in\{00,01\}}\hat{\pi}_l \pr(P_{j1},P_{j2}|Z_{jl} ; \hat \Theta),\nonumber\\
  \pr(P_{j1},P_{j2},Z_{j00}+Z_{j10}=1 ; \hat \Theta) &= \sum_{l\in\{00,10\}}\hat{\pi}_l \pr(P_{j1},P_{j2}|Z_{jl} ; \hat \Theta),
\end{align}
and $\{\hat{\pi}_{l}\}_{l\in\{00,10,01,11\}}$ and $\hat \Theta$ are estimated parameters from the GPA model.

When annotation data are available, the false discovery rates can be calculated  as
\begin{align}\label{}
  &\mbox{fdr}_1(P_{j1},P_{j2},\bfA ; \hat \Theta)=\hat \pr(Z_{j00}+Z_{j01}=1|P_{j1},P_{j2},\bfA)=\frac{\pr(P_{j1},P_{j2},\bfA,Z_{j00}+Z_{j01}=1 ; \hat \Theta)}{\pr(P_{j1},P_{j2},\bfA ; \hat \Theta)},\nonumber\\
  &\mbox{fdr}_2(P_{j1},P_{j2},\bfA ; \hat \Theta)=\hat \pr(Z_{j00}+Z_{j10}=1|P_{j1},P_{j2},\bfA)=\frac{\pr(P_{j1},P_{j2},\bfA,Z_{j00}+Z_{j10}=1 ; \hat \Theta)}{\pr(P_{j1},P_{j2},\bfA ; \hat \Theta)},
\end{align}
where
\begin{align}\label{}
  \pr(P_{j1},P_{j2},\bfA ; \hat \Theta) &= \sum_{l\in\{00,10,01,11\}} \left(\hat{\pi}_l\pr(P_{j1},P_{j2}|Z_{jl} ; \hat \Theta)\prod^D_{d=1}\pr(A_{jd}|Z_{jl}) ; \hat \Theta\right),\nonumber\\
  \pr(P_{j1},P_{j2},\bfA,Z_{j00}+Z_{j01}=1 ; \hat \Theta) &= \sum_{l\in\{00,01\}} \left(\hat{\pi}_l\pr(P_{j1},P_{j2}|Z_{jl} ; \hat \Theta)\prod^D_{d=1}\pr(A_{jd}|Z_{jl} ; \hat \Theta)\right),\nonumber\\
  \pr(P_{j1},P_{j2},\bfA,Z_{j00}+Z_{j10}=1 ; \hat \Theta) &= \sum_{l\in\{00,10\}} \left(\hat{\pi}_l \pr(P_{j1},P_{j2}|Z_{jl} ; \hat \Theta)\prod^D_{d=1}\pr(A_{jd}|Z_{jl} ; \hat \Theta)\right).
\end{align}
Finally, we use the \textit{direct posterior probability approach} \cite{newton2004} to control global false discovery rates to determine associated SNPs. Specifically, given the GPA model fitting, we first sort SNPs by their local false discovery rates from the smallest one to the largest one. Let's denote local false discovery rates of these sorted SNPs as $f_j$. We increase the threshold for local false discovery rates, $\kappa$, from zero to one until
\begin{align}\label{}
Fdr = \frac{ \sum_{j=1}^J f_j 1\left\{ f_j \leq \kappa \right\} }{ \sum_{j=1}^J 1\left\{ f_j \leq \kappa \right\} } \leq \tau,
\end{align}
where $\tau$ is the pre-determined bound of global false discovery rates. Finally, we determine SNPs with corresponding $f_j < \kappa$ to be associated with the phenotype.

\subsection*{Estimation of standard error}
To understand the genetic architecture of complex traits, we are interested in the accuracy of the estimated parameters ($\{\pi_l\},\{q_{d,l}\}$) from the GPA model.
Here we consider the standard errors of those parameters which can be calculated from covariance matrix estimated using the empirical observed information matrix \cite{mclachlan2008}. Specifically, the empirical observed information matrix is defined as
\begin{eqnarray*}
I_e ( \hat{ \Theta } ; \mathbf{P}, \mathbf{A} ) = \sum_{j=1}^M s_j ( \mathbf{P}_j, \mathbf{A}_j ; \hat{ \Theta } ) s_j^T ( \mathbf{P}_j, \mathbf{A}_j ; \hat{ \Theta } ),
\end{eqnarray*}
where $\hat{ \Theta }$ are the parameter estimates when the EM algorithm converges (i.e., maximum likelihood estimator (MLE)),
\begin{eqnarray*}
s_j ( \mathbf{P}_j, \mathbf{A}_j ; \Theta ) = E_{ \Theta } \left\{ \frac{ \partial{ \log L_{cj} ( \Theta ) } }{ \partial{ \Theta } } | \mathbf{P}, \mathbf{A} \right\},
\end{eqnarray*}
and $\log L_{cj} ( \Theta )$ is complete log likelihood for $j$-th SNP. $s_j ( \mathbf{P}_j, \mathbf{A}_j ; \Theta )$ is a column vector of length $(2^K-1 + K + 2^K D)$. The first three ($2^K-1$) components of $s_j$ correspond to $\pi_{10}, \pi_{01}, \pi_{11}$:
\begin{eqnarray*}
s_j^{(1)} ( \mathbf{P}_j, \mathbf{A}_j ; \hat{ \Theta } ) = \frac{ \hat z_{j10} }{ \hat \pi_{10} } - \frac{ \hat z_{j00} }{ \hat \pi_{00} },
s_j^{(2)} ( \mathbf{P}_j, \mathbf{A}_j ; \hat{ \Theta } ) = \frac{ \hat z_{j01} }{ \hat \pi_{01} } - \frac{ \hat z_{j00} }{ \hat \pi_{00} },
s_j^{(3)} ( \mathbf{P}_j, \mathbf{A}_j ; \hat{ \Theta } ) = \frac{ \hat z_{j11} }{ \hat \pi_{11} } - \frac{ \hat z_{j00} }{ \hat \pi_{00} },
\end{eqnarray*}
where $\hat z_{jl} = P\left( Z_{jl} = 1 | \mathbf{P}, \mathbf{A}, \hat \Theta \right), l\in\{00,10,01,11\}$. The next two ($K$) components of $s_j$ correspond to $\alpha_1, \alpha_2$:
\begin{eqnarray*}
s_j^{(4)} ( \mathbf{P}_j, \mathbf{A}_j ; \hat{ \Theta } ) = ( \hat z_{j10} + \hat z_{j11} ) \left\{ \log P_{j1} + 1 / \hat \alpha_1 \right\},\\
s_j^{(5)} ( \mathbf{P}_j, \mathbf{A}_j ; \hat{ \Theta } ) = ( \hat z_{j01} + \hat z_{j11} ) \left\{ \log P_{j2} + 1 / \hat \alpha_2 \right\}.
\end{eqnarray*}
The remaining ($2^K D$) components of $s_j$ correspond to $q_{d00}, q_{d10}, q_{d01}, q_{d11}, d = 1, \cdots, D$:
\begin{eqnarray*}
s_j^{(6)} ( \mathbf{P}_j, \mathbf{A}_j ; \hat{ \Theta } ) = ( \hat z_{j00} )( \frac{ A_{jd} }{ \hat q_{d00} } - \frac{ 1 - A_{jd} }{ 1 - \hat q_{d00} } ),
s_j^{(7)} ( \mathbf{P}_j, \mathbf{A}_j ; \hat{ \Theta } ) = ( \hat z_{j10} )( \frac{ A_{jd} }{ \hat q_{d10} } - \frac{ 1 - A_{jd} }{ 1 - \hat q_{d10} } ),\\
s_j^{(8)} ( \mathbf{P}_j, \mathbf{A}_j ; \hat{ \Theta } ) = ( \hat z_{j01} )( \frac{ A_{jd} }{ \hat q_{d01} } - \frac{ 1 - A_{jd} }{ 1 - \hat q_{d01} } ),
s_j^{(9)} ( \mathbf{P}_j, \mathbf{A}_j ; \hat{ \Theta } ) = ( \hat z_{j11} )( \frac{ A_{jd} }{ \hat q_{d11} } - \frac{ 1 - A_{jd} }{ 1 - \hat q_{d11} } ).
\end{eqnarray*}
The last part is simply ignored when annotation data is not incorporated.

We do not include a component corresponding to $\pi_{00}$ in $s_j ( \mathbf{P}_j, \mathbf{A}_j ; \Theta )$ because of the relationship that $\pi_{00} = 1 - \pi_{10} - \pi_{01}- \pi_{11}$. Instead, after we estimate the empirical observed information matrix, we estimate the standard error for $\pi_{00}$ using the Delta method \cite{shao2003} as
\begin{eqnarray*}
se( \hat \pi_{00} ) = \sqrt{ \left\{ g \prime ( \Theta ) \right\}^T \left\{ I_e ( \hat{ \Theta } ; \mathbf{P}, \mathbf{A} ) \right\}^{ -1 } \left\{ g \prime ( \Theta ) \right\} },
\end{eqnarray*}
where $g ( \Theta ) = 1 - \pi_{10} - \pi_{01}- \pi_{11}$, $g \prime ( \Theta ) = [ -1, -1, -1, \mathbf{0} ]$, and $\mathbf{0}$ is the zero vector of length $(K + 2^K D)$.

\bibliographystyle{plainnat}
\bibliography{ref,literature}

\begin{thebibliography}{10}

\bibitem{allen2010hundreds}
Hana~Lango Allen, Karol Estrada, Guillaume Lettre, Sonja~I Berndt, Michael~N
  Weedon, Fernando Rivadeneira, and et~al.
\newblock Hundreds of variants clustered in genomic loci and biological
  pathways affect human height.
\newblock {\em Nature}, 467(7317):832--838, 2010.

\bibitem{andreassen2013improved}
Ole~A Andreassen, Srdjan Djurovic, Wesley~K Thompson, Andrew~J Schork,
  Kenneth~S Kendler, Michael~C O'Donovan, and et~al.
\newblock Improved detection of common variants associated with schizophrenia
  by leveraging pleiotropy with cardiovascular-disease risk factors.
\newblock {\em The American Journal of Human Genetics}, 92(2):97--109, 2013.

\bibitem{regulomedb}
A.P. Boyle, E.L. Hong, M.~Hariharan, Y.~Cheng, M.A. Schaub, M.~Kasowski, and
  et~al.
\newblock {Annotation of functional variation in personal genomes using
  RegulomeDB.}
\newblock {\em Genome Research}, 22(9):1790--1797, 2012.

\bibitem{cantor2010}
R.M. Cantor, K.~Lange, and J.S. Sinsheimer.
\newblock {Prioritizing GWAS results: A review of statistical methods and
  recommendations for their application}.
\newblock {\em The American Journal of Human Genetics}, 86(1):6--22, 2010.

\bibitem{pgc2013genetic}
{Cross-Disorder Group of the Psychiatric Genomics Consortium}.
\newblock {Genetic relationship between five psychiatric disorders estimated
  from genome-wide SNPs}.
\newblock {\em Nature genetics}, 45(9):984--994, 2013.

\bibitem{pgp2013improved}
{Cross-Disorder Group of the Psychiatric Genomics Consortium}.
\newblock Identification of risk loci with shared effects on five major
  psychiatric disorders: a genome-wide analysis.
\newblock {\em Lancet}, 381(9875):1371--1379, 2013.

\bibitem{dempster1977}
Arthur~P Dempster, Nan~M Laird, and Donald~B Rubin.
\newblock Maximum likelihood from incomplete data via the {EM} algorithm.
\newblock {\em Journal of the Royal Statistical Society. Series B
  (Methodological)}, pages 1--38, 1977.

\bibitem{edwards2013beyond}
Stacey~L Edwards, Jonathan Beesley, Juliet~D French, and Alison~M Dunning.
\newblock {Beyond GWASs: Illuminating the Dark Road from Association to
  Function}.
\newblock {\em The American Journal of Human Genetics}, 93(5):779--797, 2013.

\bibitem{efron2010large}
B.~Efron.
\newblock {\em {Large-Scale Inference: Empirical Bayes Methods for Estimation,
  Testing, and Prediction}}.
\newblock Cambridge University Press, 2010.

\bibitem{efron2008microarrays}
Bradley Efron.
\newblock {M}icroarrays, empirical {B}ayes and the two-groups model.
\newblock {\em Statistical Science}, pages 1--22, 2008.

\bibitem{hindorff2009potential}
L.A. Hindorff, P.~Sethupathy, H.A. Junkins, E.M. Ramos, J.P. Mehta, F.S.
  Collins, and T.A. Manolio.
\newblock {Potential etiologic and functional implications of genome-wide
  association loci for human diseases and traits}.
\newblock {\em Proceedings of the National Academy of Sciences}, 106(23):9362,
  2009.

\bibitem{lee2012estimating}
S~Hong Lee, Teresa~R DeCandia, Stephan Ripke, Jian Yang, Patrick~F Sullivan,
  Michael~E Goddard, and et~al.
\newblock {Estimating the proportion of variation in susceptibility to
  schizophrenia captured by common SNPs}.
\newblock {\em Nature genetics}, 44(3):247--250, 2012.

\bibitem{lee2011estimating}
Sang~Hong Lee, Naomi~R Wray, Michael~E Goddard, and Peter~M Visscher.
\newblock Estimating missing heritability for disease from genome-wide
  association studies.
\newblock {\em The American Journal of Human Genetics}, 88(3):294--305, 2011.

\bibitem{lee2012estimation}
Sang~Hong Lee, Jian Yang, Michael~E Goddard, Peter~M Visscher, and Naomi~R
  Wray.
\newblock Estimation of pleiotropy between complex diseases using
  single-nucleotide polymorphism-derived genomic relationships and restricted
  maximum likelihood.
\newblock {\em Bioinformatics}, 28(19):2540--2542, 2012.

\bibitem{li2013improving}
Cong Li, Can Yang, Joel Gelernter, and Hongyu Zhao.
\newblock Improving genetic risk prediction by leveraging pleiotropy.
\newblock {\em arXiv preprint arXiv:1304.7417}, 2013.

\bibitem{maher2008personal}
B.~Maher.
\newblock {Personal genomes: The case of the missing heritability.}
\newblock {\em Nature}, 456(7218):18--21, 2008.

\bibitem{Manolio2010}
T.A. Manolio.
\newblock {Genomewide association studies and assessment of the risk of
  disease}.
\newblock {\em The New England Journal of Medicine}, 363(2):166--176, 2010.

\bibitem{manolio2009finding}
Teri~A Manolio, Francis~S Collins, Nancy~J Cox, David~B Goldstein, Lucia~A
  Hindorff, David~J Hunter, and et~al.
\newblock Finding the missing heritability of complex diseases.
\newblock {\em Nature}, 461(7265):747--753, 2009.

\bibitem{mclachlan2008}
Geoffrey McLachlan and Thriyambakam Krishnan.
\newblock {\em The {EM} algorithm and extensions}.
\newblock John Wiley \& Sons, 2008.

\bibitem{morris2012large}
Andrew~P Morris, Benjamin~F Voight, Tanya~M Teslovich, Teresa Ferreira,
  Ayellet~V Segre, Valgerdur Steinthorsdottir, and et~al.
\newblock Large-scale association analysis provides insights into the genetic
  architecture and pathophysiology of type 2 diabetes.
\newblock {\em Nature genetics}, 44(9):981--990, 2012.

\bibitem{newton2004}
Michael Newton, Amine Noueiry, Deepayan Sarkar, and Paul Ahlquist.
\newblock Detecting differential gene expression with a semiparametric
  hierarchical mixture method.
\newblock {\em Biostatistics}, 5(2):155--176, 2004.

\bibitem{nicolae2010trait}
Dan~L Nicolae, Eric Gamazon, Wei Zhang, Shiwei Duan, M~Eileen Dolan, and
  Nancy~J Cox.
\newblock {Trait-associated SNPs are more likely to be eQTLs: annotation to
  enhance discovery from GWAS}.
\newblock {\em PLoS genetics}, 6(4):e1000888, 2010.

\bibitem{pounds2003}
Stan Pounds and Stephan~W Morris.
\newblock Estimating the occurrence of false positives and false negatives in
  microarray studies by approximating and partitioning the empirical
  distribution of p-values.
\newblock {\em Bioinformatics}, 19(10):1236--1242, 2003.

\bibitem{raychaudhuri2010accurately}
Soumya Raychaudhuri, Joshua~M Korn, Steven~A McCarroll, David Altshuler, Pamela
  Sklar, Shaun Purcell, Mark~J Daly, et~al.
\newblock Accurately assessing the risk of schizophrenia conferred by rare
  copy-number variation affecting genes with brain function.
\newblock {\em PLoS genetics}, 6(9):e1001097, 2010.

\bibitem{rothman2010multi}
Nathaniel Rothman, Montserrat Garcia-Closas, Nilanjan Chatterjee, Nuria Malats,
  Xifeng Wu, Jonine~D Figueroa, Francisco~X Real, David Van Den~Berg, Giuseppe
  Matullo, Dalsu Baris, et~al.
\newblock A multi-stage genome-wide association study of bladder cancer
  identifies multiple susceptibility loci.
\newblock {\em Nature genetics}, 42(11):978--984, 2010.

\bibitem{sakoda2013turning}
Lori~C Sakoda, Eric Jorgenson, and John~S Witte.
\newblock {Turning of COGS moves forward findings for hormonally mediated
  cancers}.
\newblock {\em Nature Genetics}, 45(4):345--348, 2013.

\bibitem{schork2013all}
Andrew~J Schork, Wesley~K Thompson, Phillip Pham, Ali Torkamani, J~Cooper
  Roddey, Patrick~F Sullivan, John~R Kelsoe, Michael~C O'Donovan, Helena
  Furberg, Nicholas~J Schork, et~al.
\newblock {All SNPs are not created equal: genome-wide association studies
  reveal a consistent pattern of enrichment among functionally annotated SNPs}.
\newblock {\em PLoS genetics}, 9(4):e1003449, 2013.

\bibitem{shao2003}
Jun Shao.
\newblock {\em Mathematical statistics}.
\newblock Springer, 2nd edition, 2003.

\bibitem{shriner2012moving}
Daniel Shriner.
\newblock Moving toward system genetics through multiple trait analysis in
  genome-wide association studies.
\newblock {\em Frontiers in genetics}, 3, 2012.

\bibitem{sivakumaran2011abundant}
Shanya Sivakumaran, Felix Agakov, Evropi Theodoratou, James~G Prendergast, Lina
  Zgaga, Teri Manolio, and et~al.
\newblock {Abundant pleiotropy in human complex diseases and traits}.
\newblock {\em The American Journal of Human Genetics}, 89(5):607--618, 2011.

\bibitem{sklar2011large}
Pamela Sklar, Stephan Ripke, Laura~J Scott, Ole~A Andreassen, Sven Cichon, Nick
  Craddock, Howard~J Edenberg, John~I Nurnberger, Marcella Rietschel, Douglas
  Blackwood, et~al.
\newblock {Large-scale genome-wide association analysis of bipolar disorder
  identifies a new susceptibility locus near ODZ4}.
\newblock {\em Nature genetics}, 43(10):977, 2011.

\bibitem{solovieff2013pleiotropy}
Nadia Solovieff, Chris Cotsapas, Phil~H Lee, Shaun~M Purcell, and Jordan~W
  Smoller.
\newblock Pleiotropy in complex traits: challenges and strategies.
\newblock {\em Nature Reviews Genetics}, 2013.

\bibitem{subramanian2005gene}
Aravind Subramanian, Pablo Tamayo, Vamsi~K Mootha, Sayan Mukherjee, Benjamin~L
  Ebert, Michael~A Gillette, Amanda Paulovich, Scott~L Pomeroy, Todd~R Golub,
  Eric~S Lander, et~al.
\newblock Gene set enrichment analysis: a knowledge-based approach for
  interpreting genome-wide expression profiles.
\newblock {\em Proceedings of the National Academy of Sciences of the United
  States of America}, 102(43):15545--15550, 2005.

\bibitem{encode2012}
{The ENCODE Project Consortium}.
\newblock {An integrated encyclopedia of DNA elements in the human genome}.
\newblock {\em Nature}, 489:57--74, 2012.

\bibitem{thurman2012accessible}
Robert~E Thurman, Eric Rynes, Richard Humbert, Jeff Vierstra, Matthew~T
  Maurano, Eric Haugen, Nathan~C Sheffield, Andrew~B Stergachis, Hao Wang,
  Benjamin Vernot, et~al.
\newblock The accessible chromatin landscape of the human genome.
\newblock {\em Nature}, 489(7414):75--82, 2012.

\bibitem{vattikuti2012heritability}
Shashaank Vattikuti, Juen Guo, and Carson~C Chow.
\newblock {Heritability and genetic correlations explained by common SNPs for
  metabolic syndrome traits}.
\newblock {\em PLoS genetics}, 8(3):e1002637, 2012.

\bibitem{visscher2008sizing}
Peter~M Visscher.
\newblock Sizing up human height variation.
\newblock {\em Nature genetics}, 40(5):489--490, 2008.

\bibitem{visscher2012five}
Peter~M Visscher, Matthew~A Brown, Mark~I McCarthy, and Jian Yang.
\newblock {Five years of GWAS discovery}.
\newblock {\em The American Journal of Human Genetics}, 90(1):7--24, 2012.

\bibitem{visscher2008heritability}
Peter~M Visscher, William~G Hill, and Naomi~R Wray.
\newblock Heritability in the genomics era - concepts and misconceptions.
\newblock {\em Nature Reviews Genetics}, 9(4):255--266, 2008.

\bibitem{noncoding2012}
L.D. Ward and M.~Kellis.
\newblock {Interpreting noncoding genetic variation in complex traits and human
  disease}.
\newblock {\em Nature Biotechnology}, 30:1095--1106, 2012.

\bibitem{yang2010common}
Jian Yang, Beben Benyamin, Brian~P McEvoy, Scott Gordon, Anjali~K Henders,
  Dale~R Nyholt, and et~al.
\newblock {Common SNPs explain a large proportion of the heritability for human
  height}.
\newblock {\em Nature genetics}, 42(7):565--569, 2010.

\bibitem{yang2011gcta}
Jian Yang, S~Hong Lee, Michael~E Goddard, and Peter~M Visscher.
\newblock {GCTA: a tool for genome-wide complex trait analysis}.
\newblock {\em The American Journal of Human Genetics}, 88(1):76--82, 2011.

\bibitem{yang2011genome}
Jian Yang, Teri~A Manolio, Louis~R Pasquale, Eric Boerwinkle, Neil Caporaso,
  Julie~M Cunningham, and et~al.
\newblock Genome partitioning of genetic variation for complex traits using
  common snps.
\newblock {\em Nature genetics}, 43(6):519--525, 2011.

\end{thebibliography}

\end{document}